\documentclass[prd,twocolumn,showpacs,amsmath,amssymb,nofootinbib,eqsecnum,floatfix]{revtex4}

\usepackage{graphicx}
\usepackage{bm}
\begin{document}
%\draft
\preprint{gr-qc/}
\title{Dynamics of spinning test particles in Kerr spacetime}
\author{Michael D. Hartl}
\email{mhartl@tapir.caltech.edu}
\affiliation{
Department of Physics, California Institute of Technology, Pasadena CA 91125}
\date{October 14, 2002}
\begin{abstract}

We investigate the dynamics of  relativistic spinning test particles in the
spacetime of a rotating black hole using the Papapetrou equations. We use the
method of Lyapunov exponents to determine whether the orbits exhibit sensitive
dependence on initial conditions, a signature of chaos. In the case of
maximally spinning equal-mass binaries (a limiting case that violates the
test-particle approximation) we find unambiguous positive Lyapunov exponents
that come in pairs~$\pm\lambda$, a characteristic of Hamiltonian dynamical
systems. We find no evidence for nonvanishing Lyapunov exponents for physically
realistic spin parameters, which suggests that chaos may not manifest itself in
the gravitational radiation of extreme mass-ratio binary black-hole inspirals
(as detectable, for example, by LISA, the Laser Interferometer Space Antenna).

\end{abstract}
\maketitle

\section{INTRODUCTION}

The presence of chaos (or lack thereof) in relativistic binary inspiral systems
has received intense attention recently due to the implications for
gravitational-wave detection~\citep{LevinPRL2000,Levin2000,SuzukiMaeda1997,
SchnittmanRasio2001,Cornish2001,CornishLevin2002,CornishLevinSR2002,Hughes2001},
especially regarding the generation of theoretical templates for use in matched
filters.  There is concern that the sensitive dependence to initial conditions
that characterizes chaos may make the calculation of such templates difficult
or impossible~\citep{Hughes2001}.  In particular, in the presence of chaos the
number of templates would increase exponentially with the number of wave cycles
to be fitted.  In addition to this important concern, the problem of chaos in
general relativity  has inherent interest, as the dynamical behavior of general
relativistic systems is poorly understood.

Several authors have reported the presence of chaos for systems of two point
masses in which one or both particles are spinning~\citep{SuzukiMaeda1997,
LevinPRL2000, CornishLevin2002}. Our work follows up
on~\citep{SuzukiMaeda1997}, which studies the dynamics of a spinning test
particle orbiting a nonrotating (Schwarzschild) black hole using the Papapetrou
equations~[Eqs.~(\ref{eq:Papapetrou})]. We extend this work to a rotating (Kerr)
black hole, motivated by the expectation that many astrophysically relevant
black holes have nonzero angular momentum.  Furthermore, the potential for
chaos may be greater in Kerr spacetime since the Kerr metric has less symmetry
and hence fewer integrals of the motion than Schwarzschild.  In addition, the
decision to focus on test particles is motivated partially by the LISA
gravitational wave detector~\citep{LISAweb}, which will be sensitive to
radiation from spinning compact objects orbiting supermassive black holes in
galactic nuclei. Using the Kerr metric is appropriate since  such supermassive
black holes will in general have nonzero spin.

There are many techniques for investigating chaos in dynamical systems, but for
the case at hand we favor the use of Lyapunov exponents to quantify chaos. 
Informally, if~$\epsilon_0$ is the phase-space distance between two nearby
initial conditions in phase space, then for chaotic systems the separation
grows exponentially (\emph{sensitive dependence on initial conditions}):
$\epsilon(\tau)=\epsilon_0\,e^{\lambda \tau}$, where $\lambda$ is the Lyapunov
exponent.  (See Sec.~\ref{sec:lyapunov_exponents} for a discussion of issues
related to the choice of metric used to determine the distance in phase
space.)  The value of Lyapunov exponents lies not only in establishing chaos,
but also in providing a characteristic timescale $\tau_\lambda=1/\lambda$ for
the exponential separation.

By definition, chaotic orbits are bounded phase space flows with at least
one nonzero Lyapunov exponent. There are additional technical requirements
for chaos that rule out periodic or quasiperiodic orbits, equilibria, and 
other types of patterned behavior~\cite{ASY1997}.  For example, 
unstable circular orbits in Schwarzschild spacetime can have positive 
Lyapunov exponents~\cite{Cornish2001}, but such orbits are completely 
integrable (see Sec.~\ref{sec:integrability}) and hence not chaotic.
In practice, we restrict ourselves to generic orbits, avoiding the
specialized initial conditions that lead to positive Lyapunov exponents
in the absence of chaos.

The use of Lyapunov exponents is potentially dangerous in general
relativity because of the freedom to re-define the time coordinate. Chaos
can seemingly be removed by a coordinate transformation: simply let
$\tau'=\log\tau$ and the chaos disappears.  
Fortunately, in our case
there is a fixed background spacetime with a time coordinate that
is not dynamical but rather is simply a re-parameterization of the
proper time. As a result, we will not encounter this time coordinate
re-definition ambiguity (which plagued, for example, attempts to establish
chaos in mixmaster cosmological models, until coordinate-invariant
methods were developed~\citep{CornishLevin1997}). 
Furthermore, we can compare times in different coordinate systems using 
ratios: if~$t_p$ is the period of a periodic orbit in 
some coordinate system with time coordinate~$t$, and~$\tau_p$ is the 
period in proper time, then their ratio provides a conversion factor between
times in different coordinate systems~\cite{Cornish2001}:
\begin{equation}
\label{eq:tratio}
\frac{t}{\tau} = \frac{t_p}{\tau_p}.
\end{equation}
For chaotic orbits,
which are not periodic, we use the average value of $dt/d\tau$ over the
orbit, so that 
\begin{equation}
\label{eq:taut1}
\frac{t_\lambda}{\tau_\lambda} = \left\langle \frac{dt}{d\tau}\right\rangle
\end{equation}
as discussed in Sec.~\ref{sec:comments}.  [This more general
formula reduces to Eq.~(\ref{eq:tratio}) in the case of periodic orbits.]
Since we want to
measure the local divergence of trajectories, the natural definition is
to use the divergence in local Lorentz frames, which suggests that we
use the proper time~$\tau$ as our time parameter.  The Lyapunov timescale
in any coordinate system can then be obtained using Eq.~(\ref{eq:taut1}).

Lyapunov exponents provide a quantitative definition of chaos, but there are
several common qualitative methods as well, none of which we use in the present
case, for reasons explained below.   Perhaps the most common qualitative tool
in the analysis of dynamical systems is the use of Poincar\'{e} surfaces of
section.  Poincar\'{e} sections reduce the phase space by one dimension by
considering the intersection of the phase space trajectory with some fixed
surface, typically taken to be a plane. Plotting momentum vs.\ position for
intersections of the trajectory with this  surface then gives a qualitative
view of the dynamics. As noted in~\citep{SchnittmanRasio2001}, such sections
are most useful when the number of degrees of freedom minus the number of
constraints (including integrals of the motion) is not greater than two, since
in this case the resulting points fall on a one-dimensional curve for
non-chaotic orbits, but are ``dusty'' for chaotic orbits (and in the case of
dissipative dynamical systems lie on fractal attractors).  Unfortunately, the
system we consider has too many degrees of freedom for Poincar\'{e} sections to
be useful. It is possible to plot momentum vs.\ position when the trajectory
intersects a section that is a plane in physical space (say
$x=0$)~\citep{SuzukiMaeda1997}, but this is not in general a true Poincar\'{e}
section.\footnote{In~\citep{SuzukiMaeda1997}, they are aided by the symmetry of
Schwarzschild spacetime, which guarantees that one component of the spin
tensor~(Sec.~\ref{sec:Papapetrou_eq} below)  is zero in the equatorial plane. 
As a result, it turns out that all but two of their variables are determined on
the surface, and thus their sections are valid. Unfortunately, the reduced
symmetry of the Kerr metric makes this method unsuitable for the system we
consider in this paper.}

Other qualitative methods include power spectra and chaotic attractors.
The power spectra for regular orbits have a finite number of discrete
frequencies, whereas their chaotic counterparts are continuous.
Unfortunately, it is difficult to differentiate between complicated
regular orbits, quasiperiodic orbits, and chaotic orbits, so we
have avoided their use.  Chaotic attractors, which typically involve
orbits asymptotically attracted to a fractal structure, are powerful
tools for exploring chaos, but their use is limited to dissipative
systems~\citep{ASY1997}. Nondissipative systems, including test particles
in general relativity, do not possess attractors~\citep{Ott1993}.

Following Suzuki and Maeda~\citep{SuzukiMaeda1997}, we use the Papapetrou
equations to model the dynamics of a spinning test particle
in the absence of gravitational radiation. 
We extend their work in a Schwarzschild background
by considering orbits in Kerr spacetime, and we also improve on
their methods for calculating Lyapunov exponents.
The most significant improvement
is the use of a rigorous method for determining Lyapunov exponents using
the linearized equations of motion for each trajectory in phase 
space~(Sec.~\ref{sec:lyapunov_exponents}), which requires knowledge of the Jacobian
matrix for the Papapetrou system~(Sec.~\ref{sec:Jacobian_matrix}).
We augment this method with an implementation of an
informal deviation vector approach, which tracks the size of an initial
deviation of size~$\epsilon_0$ and uses the relation $\epsilon(\tau) = 
\epsilon_0\,e^{\lambda\tau}$ discussed above.  We are careful in all cases
to incorporate the constrained nature of the Papapetrou equations
(Sec.~\ref{sec:Papapetrou_eq})
in the calculation of Lyapunov exponents (Sec.~\ref{sec:constraint_complexities}).

We use units where $G=c=1$ and sign conventions as in MTW~\citep{MTW}. We
use vector arrows for 4-vectors (e.g., $\vec p$ for the 4-momentum)
and boldface for Euclidean vectors (e.g., $\bm{\xi}$ for a Euclidean
tangent vector). The symbol~$\log$ refers in all cases to the natural
logarithm $\log_e$.

\section{Spinning test particles}

\subsection{Papapetrou equations}
\label{sec:Papapetrou_eq}

The Papapetrou equations~\citep{Papapetrou} describe the motion of
a spinning test particle.
Although Papapetrou first derived the equations of motion for such a
particle, the formulation by Dixon~\citep{Dixon} is the starting point for most
investigations because of its conceptual clarity.  Dixon writes the equations
of motion in terms of the 4-momentum~$p^\alpha$ and spin
tensor~$S^{\alpha\beta}$, which are defined by
integrals of the particle's stress-energy tensor~$T^{\alpha\beta}$
over an arbitrary
spacelike hypersurface~$\Sigma$:
\begin{equation}
\label{eq:4-momentum}
p^{\alpha}(\Sigma) = \int_\Sigma T^{\alpha\beta}\,d\Sigma_\beta
\end{equation} 
\begin{equation}
\label{eq:spin}
S^{\alpha\beta}(\vec z, \Sigma) = 2\int_\Sigma 
	(x^{[\alpha} - z^{[\alpha})T^{\beta]\gamma}\,d\Sigma_\gamma,
\end{equation} 
where~$\vec z$ is the coordinate of the center of mass.  The equations
of motion for a spinning test particle are then
\begin{eqnarray}
\label{eq:Dixon}
\frac{dx^\mu}{d\tau} & = & v^\mu\nonumber\\
\nabla_{\vec v}\,p^\mu&=&
	-\textstyle{\frac{1}{2}}R^{\mu}_{\ \nu\alpha\beta}\,
	v^\nu S^{\alpha\beta}\\
\nabla_{\vec v}\,S^{\mu\nu}&=&2p^{[\mu}v^{\nu]}\nonumber,  
\end{eqnarray}
where~$v^\mu$ is the 4-velocity, i.e., the 
tangent to the particle's worldline.  It is apparent that 
the 4-momentum deviates from geodesic motion due to a coupling of the spin to
the Riemann curvature.

\subsubsection{Spin supplementary conditions}

As written, the Papapetrou equations~\ref{eq:Dixon} are under-determined,
and require a \emph{spin supplementary condition}
to determine the rest frame of the particle's center of mass. 
Following Dixon, we choose 
\begin{equation}
\label{eq:p_ssc}
p_\mu S^{\mu\nu} = 0,
\end{equation}
which picks out a unique worldline that we identify as the center of mass.
In particular, in the zero
3-momentum frame defined by~$p^i = 0$,
applying Eq.~(\ref{eq:p_ssc}) to Eq.~(\ref{eq:spin})  yields
\begin{equation}
z^i  = 
	\frac{\int_{t=\textrm{const.}} x^i\,T^{00}\,d^3x}
	{\int_{t=\textrm{const.}} T^{00}\,d^3x},
\end{equation}
which is the proper relativistic generalization of the Newtonian 
center of mass.  The frame defined by~$p^i=0$ is thus the rest frame of the
center of mass, and in this frame Eq.~(\ref{eq:p_ssc}) implies that~$S^{0j}=0$,
i.e., the spin is purely spatial in the rest frame.

A second possibility for the supplementary condition is 
\begin{equation}
\label{eq:v_ssc}
v_\mu S^{\mu\nu} = 0.
\end{equation}
This condition
has the disadvantage that it is satisfied by a family of helical
worldlines filling a cylinder with frame-dependent
radius~\cite{Dixon,Moller}, centered on the worldline picked out by
condition~\ref{eq:p_ssc}.  As a result, we adopt $p_\mu S^{\mu\nu}=0$ as the
supplementary condition.

We note that the difference between the conditions~\ref{eq:p_ssc}
and~\ref{eq:v_ssc} is third order in the
spin [which follows from Eq.~(\ref{eq:w}) below],
which means that it is negligible for physically realistic spins
(Sec.~\ref{sec:spin_param}).  In particular, the two conditions are
equivalent for post-Newtonian expansions~\cite{BarkerOConnell1974}, where
condition~\ref{eq:v_ssc} is typically employed~\cite{Kidder1995}.

\subsubsection{A reformulation of the equations}

For numerical reasons, we use a form of the equations different from
Eqs.~(\ref{eq:Dixon}).  (We discuss this and other numerical considerations in
Sec.~\ref{sec:numerical}.)  Following the appendix in~\citep{SuzukiMaeda1997},
we write the equations in terms of the momentum 1-form~$p_\mu$ and the spin
1-form $S_\mu$.\footnote{The lowered indices are motivated by the Hamiltonian
formulation for a nonspinning test particle, where it is the one-form $p_\mu$
that is canonically conjugate to $x^\mu$~\cite{MTW}.}  The system under
consideration is a spinning particle of rest mass~$\mu$ orbiting a central body
of mass~$M$; in what follows, we measure all times and lengths in terms of~$M$,
and we measure the momentum of the particle in terms of~$\mu$, so that $p_\nu
p^\nu=-1$.  In these normalized units, the equations of motion are
\begin{eqnarray}
\label{eq:Papapetrou}
\frac{dx^\mu}{d\tau}&=&v^\mu\nonumber\\ \nabla_{\vec v}\,p_\mu&=&-R^{*\
\alpha\beta}_{\mu\nu}v^\nu p_\alpha S_\beta\\ \nabla_{\vec
v}\,S_\mu&=&-p_\mu \left(R^{*\alpha\ \gamma\delta}_{\ \ \beta}
	S_\alpha v^\beta p_\gamma S_\delta\right)\nonumber
\end{eqnarray}
where \begin{equation} R^{*\alpha\ \mu\nu}_{\
\ \beta}=\textstyle{{1\over2}}R^{\alpha}_{\ \beta\rho\sigma}
\epsilon^{\rho\sigma\mu\nu}.  
\end{equation}

The tensor and vector formulations of the spin are related by 
\begin{equation}
\label{eq:S_vector}
S_{\mu}=\textstyle{\frac{1}{2}}\epsilon_{\mu\nu\alpha\beta}\,
	u^\nu S^{\alpha\beta}
\end{equation}
and 
\begin{equation}
\label{eq:S_tensor}
S^{\mu\nu}=
-\epsilon^{\mu\nu\alpha\beta}S_\alpha u_\beta,
\end{equation}
where $u_\nu = p_\nu/\mu$ ($= p_\nu$ in normalized units).
In addition, the spin satisfies the
condition 
\begin{equation} S_\mu S^\mu=\textstyle{\frac{1}{2}}
	S_{\mu\nu}\,S^{\mu\nu} = S^2, 
\end{equation} 
where~$S$ is
the spin of the particle measured in units of~$\mu M$ 
(see Sec.~\ref{sec:spin_param}).  

Because of the coupling of the spin to the Riemann curvature,
the 4-momentum~$p^\mu$ [Eq.~(\ref{eq:4-momentum})] is not parallel
to the tangent~$v^\mu$.
The supplementary condition~\ref{eq:p_ssc} allows for an 
explicit solution for the difference between them
(see~\citep{Semerak1999} for a derivation):
\begin{equation}
\label{eq:v}
v^\mu=N(p^\mu+w^\mu),
\end{equation}
where
\begin{equation}
\label{eq:w}
w^\mu=-{^*}R^{*\mu\alpha\beta\gamma} S_\alpha p_\beta S_\gamma
\end{equation}
and
\begin{equation}
{^*}R^{*\alpha\beta\mu\nu}=\textstyle{{1\over2}}R^{*\alpha\beta\rho\sigma}
\epsilon_{\rho\sigma}^{\ \ \ \mu\nu}.
\end{equation}
The normalization constant~$N$ is fixed by the constraint $v_\mu v^\mu=-1$.
We see from Eq.~(\ref{eq:w}) that the difference between~$p^\mu$ and~$v^\mu$
is~${\cal O}(S^2)$, so that the difference between Eqs.~(\ref{eq:p_ssc})
and~\ref{eq:v_ssc} is~${\cal O}(S^3)$.

The spin 1-form satisfies two orthogonality constraints:
\begin{equation}
\label{eq:constraint_1}
p^\mu S_\mu = 0
\end{equation}
and
\begin{equation}
\label{eq:constraint_2}
v^\mu S_\mu = 0.
\end{equation}
These two constraints are equivalent as long as~$v^\mu$ is given by 
Eq.~(\ref{eq:v}), since~$w^\mu S_\mu \propto  {^*}R^{*\mu\alpha\beta\gamma} S_\mu
S_\alpha \equiv 0$.  When parameterizing the initial conditions, we enforce
Eq.~(\ref{eq:constraint_1}); since we use  Eq.~(\ref{eq:v}) in the equations of
motion, Eq.~(\ref{eq:constraint_2}) is then automatically satisfied.

\subsubsection{Range of validity}

We note that the Papapetrou equations include effects due only to the mass
monopole and spin dipole (the pole-dipole approximation). In particular, the
tidal coupling, which is a mass quadrupole effect, is neglected. It is also
important to note that the Papapetrou equations are conservative and hence
ignore the effects of gravitational radiation.  For a thorough and accessible
general discussion of the Papapetrou equations and related matters, including a
comprehensive literature review, see Semer\'{a}k~\citep{Semerak1999}.

\subsection{Comments on the spin parameter}
\label{sec:spin_param}

It is crucial to note that, in our normalized
units, the spin parameter~$S$ is measured in terms of~$\mu M$, not~$\mu^2$.
The system we consider in this paper
is a compact spinning body of mass~$\mu$ orbiting a large
body of mass~$M$, which we take to be a supermassive
Kerr black hole satisfying~$M\approx10^5$--$10^6\,M_\odot$.
We will show that
physically realistic values of the spin must satisfy~$S\ll1$ for the
compact objects (black holes, neutron stars,
and white dwarfs) most relevant for the test particles
described by the Papapetrou equations.\footnote{Recall
that the Papapetrou equations ignore tidal coupling, so they are inappropriate
for modeling more extended objects.}
The case of a black hole is simplest:
a maximally spinning black hole of mass~$\mu$ has spin angular 
momentum~$s=\mu^2$, so a small black hole~$\mu$ orbiting a large black hole of 
mass~$M\gg\mu$ has a small spin parameter~$S$:
\[ S = \frac{s}{\mu M}\leq\frac{\mu^2}{\mu M}=
\frac{\mu}{M}\ll1.\]
The limit is similar for neutron stars:
most models of neutron stars have a maximum spin of 
$s_{\mathrm{max}}\approx0.6\,\mu^2$~\citep{CookShapiroTeukolsky1994}, which
gives $S\alt0.6\,\mu/M$. 

\subsubsection{Bounds on $S$ for stellar objects}

The bound on~$S$ is relatively simple for black holes and neutron stars, 
but the situation is more complicated for compact stellar objects such
as white dwarfs.
The maximum spin of a stellar object
is typically determined by the mass-shedding limit, i.e.,
the maximum spin before the star begins to break up.
The spin in the case of the break-up limit is the moment of inertia
times the maximum (break-up) angular velocity:
$s_\mathrm{max} = I\Omega_\mathrm{max}$.  If we write $I = \alpha \mu R^2$
and $\Omega_\mathrm{max} = \beta\sqrt{G\mu/R^3}$ for some 
constants~$\alpha$,~$\beta\alt1$, then we have 
\begin{equation}
\label{eq:s_max}
s_\mathrm{max} = \alpha\beta\,(G\mu^3 R)^{1/2}.
\end{equation}
The values of $\alpha$~and~$\beta$ depend on the
stellar model; if we use the values for an~$n=1.5$ 
polytrope, we get $\alpha = 0.2044$ and $\beta = 0.5366$~\citep{LaiRasioShapiro1993},
so that
$s_\mathrm{max} = 0.110\,(G\mu^3 R)^{1/2}$.  

The limit in~Eq.~(\ref{eq:s_max})
depends on the mass-radius relation for the object in question.
Since most neutron stars have masses and radii in a narrow range, 
the estimate of $s_{\mathrm{max}}\approx0.6\,\mu^2$ discussed above is sufficient, 
but for white dwarfs the value of~$s_{\mathrm{max}}$ can depend
strongly on the mass.  
An analytical approximation for the mass-radius
relation for non-rotating white dwarfs
is~\citep{Nauenberg1972}:\footnote{The mean molecular weight~$\bar\mu$
is set equal to~$2$, corresponding to helium and heavier elements,
which is appropriate for most astrophysical white dwarfs.}
\begin{equation}
\label{eq:wd_R}
\frac{R}{R_\odot} = 0.01125\left(\frac{\mu}{\mu_\mathrm{max}}\right)^{-1/3}\,f(\mu)^{1/2}
\end{equation}
where
\begin{equation}
	f(\mu) = 1-\left(\frac{\mu}{\mu_\mathrm{max}}\right)^{4/3}
\end{equation}
and
\begin{equation}
\mu_\mathrm{max} = 1.454\,M_\odot.
\end{equation}

\begin{figure}
\includegraphics[width=3.in]{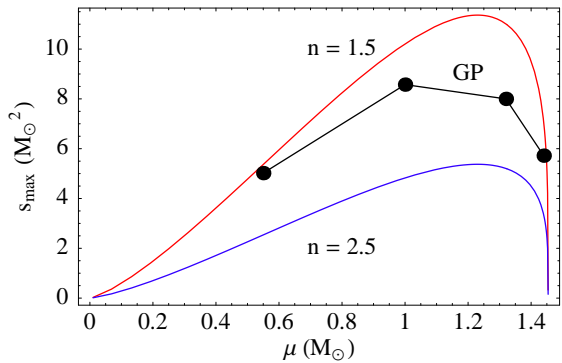}
\caption{The maximum spin angular momentum~$s_\mathrm{max}$ vs.~mass~$\mu$ for a
rigidly rotating white
dwarf.  We plot curves for $n=1.5$ and $n=2.5$ polytropic approximations
using~Eq.~(\ref{eq:s_wd}), together with four points derived
using a more realistic numerical white dwarf model
(Geroyannis~and~Papasotiriou~\citep{GP00}).}
\label{fig:s_max}
\end{figure}

We could plug Eq.~(\ref{eq:wd_R}) into~Eq.~(\ref{eq:s_max}) to obtain an
order-of-magnitude estimate, but~\citep{LaiRasioShapiro1993} tabulates a
constant~$\bar J$ equivalent to the product~$\alpha\beta$ (which increases as
the angular velocity of the star increases). They write $J = \bar J\,(GM^3
R_0)^{1/2}$ for a rotating white dwarf, where $\bar J$ depends on the
polytropic index~$n$ of a \emph{non-spinning} white dwarf of the same mass,
and~$R_0$ is the non-spinning radius.  In our notation, this reads
\begin{equation}
\label{eq:s_max_good}
s_\mathrm{max} = \bar J\,(G\mu^3 R)^{1/2}.
\end{equation}
White dwarfs with~$\mu > 0.6\,M_\odot$ are not well approximated by polytropes
(the effective polytropic index varies from near~3 in the core to near~1.5 in
the outer parts), but useful bounds can be obtained by substituting~$R$ from
Eq.~(\ref{eq:wd_R}), which is more accurate for white dwarfs than a 
pure polytrope model.
Plugging~\ref{eq:wd_R} into~\ref{eq:s_max_good}
and converting to geometric units gives
\begin{equation}
\label{eq:s_wd}
s_\mathrm{max} = 77.68\,\bar J\,\mu^{4/3}\,M_\odot^{2/3}\,f(\mu)^{1/4}.
\end{equation}
From Table~3 in~\citep{LaiRasioShapiro1993}, we have $\bar J = 0.1660$
for a maximally rotating $n=1.5$~polytrope (vs.\ $\alpha\beta = 0.110$
for a slowly rotating one)
and $\bar J = 0.0785$ for~$n=2.5$.  
As illustrated in~Fig.~\ref{fig:s_max},
the values for a more realistic numerical model~\citep{GP00}
lie between these curves, as expected.

\begin{figure}
\includegraphics[width=3.in]{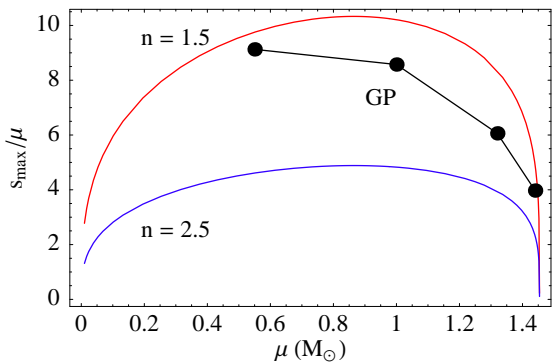}
\caption{$s_\mathrm{max}/\mu$ vs.~mass for a white
dwarf.  As in Fig.~\ref{fig:s_max}, we plot curves for $n=1.5$ and $n=2.5$ polytropes
and the numerical model from~\citep{GP00}. The corresponding spin parameter~$S_\mathrm{max}$
is obtained simply by dividing $s_\mathrm{max}/\mu$ 
by the mass~$M$ of the central black hole.}
\label{fig:S_max}
\end{figure}

Note from Eq.~(\ref{eq:s_wd}) that $s_\mathrm{max}/\mu^2\propto\mu^{-2/3}$ for
$\mu\ll\mu_\mathrm{max}$,
so that the spin per unit mass squared is unbounded 
as $\mu\rightarrow0$.\footnote{Eq.~(\ref{eq:s_wd}) is valid only for $\mu\agt 0.01\,M_\odot$,
but~$s_\mathrm{max}/\mu^2$ continues to increase with decreasing~$\mu$
for equations of state appropriate for brown dwarfs and planets.}
Nevertheless, the spin parameter~$S_\mathrm{max}$ is bounded, since 
$S_\mathrm{max} \propto s_\mathrm{max}/\mu\propto\mu^{1/3}$ in the low mass limit.
We plot~$s_\mathrm{max}/\mu$ vs.~$\mu$ in Fig.~\ref{fig:S_max}, which shows that the maximum
value of $s_\mathrm{max}/\mu$ is approximately $9\,M_\odot$ (corresponding to a 
$\mu=0.5\,M_\odot$ white dwarf).  
For a central black hole of mass~$M=10^6\,M_\odot$, we then have
\begin{equation}
S\leq S_\mathrm{max} = \frac{s_\mathrm{max}}{\mu M}
= 9\times10^{-6},
\end{equation}
which is small compared to unity.

\subsubsection{Tidal disruption}

\begin{figure}
\includegraphics[width=3.in]{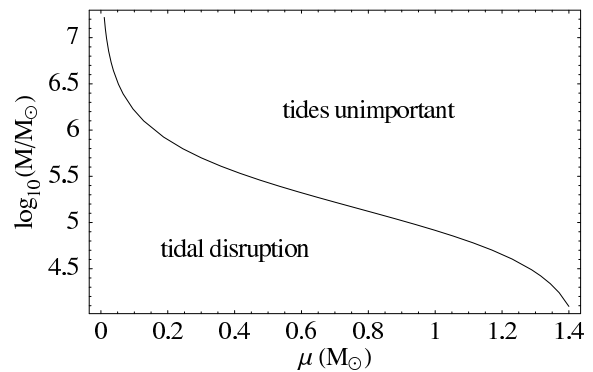}
\caption{The minimum black hole mass~$M$ required not to disrupt an inspiraling corotating
white dwarf before the last stable (prograde) circular orbit 
around a maximally rotating Kerr black hole,
as a function of white dwarf mass~$\mu$.} 
\label{fig:disrupt}
\end{figure}

We can obtain a higher value of~$S$ if the central black hole mass is smaller, but
it is important to bear in mind that such lower-mass black holes may tidally disrupt
the white dwarf companion, thereby violating a necessary condition for the validity
of the Papapetrou equations.
In order of magnitude, a white dwarf orbiting at
radius~$r$ will be disrupted
when the tidal acceleration due to the central body overcomes its self-gravity, i.e.,
\begin{equation}
\frac{GM}{r^3}R \geq \frac{G\mu}{R^2}.
\end{equation}
For the white dwarf to be undisrupted down to the horizon at~$r=M$, we must
have $M \leq R^{3/2}\mu^{-1/2}$, so that [using Eq.~(\ref{eq:wd_R})] the minimum
mass not to disrupt is
$M_\mathrm{min}\propto\mu^{-1}$.
We could evaluate the proportionality constant using Eq.~(\ref{eq:wd_R}), but
we can obtain a more accurate result by adopting a constant based
on a more realistic tidal disruption model.  Tables 
1~and~2 of~\cite{WigginsLai2000} give the value of the 
variable~$\hat r\equiv\displaystyle{\frac{r}{R}
\left(\frac{\mu}{M}\right)^{1/3}}$, which is approximately~$2.0$
for the white dwarfs of interest here.  This gives
\begin{equation}
M_\mathrm{min} = 2.0^{-3/2}\,R^{3/2}\mu^{-1/2},
\end{equation}
as illustrated in Fig.~\ref{fig:disrupt}.  For a~$1.0\,M_\odot$
white dwarf, which (based on~\cite{GP00}) has~$s_\mathrm{max} = 8.57\,M^2_\odot$,
the central black hole must satisfy~$M_\mathrm{min} = 8.2\times10^{4}\,M_\odot$, 
so that the spin parameter~$S$ can be no bigger than 
$S_\mathrm{max}=s_\mathrm{max}/(\mu M_\mathrm{min}) = 1.0\times10^{-4}$
in order to avoid tidal disruption.

\subsubsection{The $S=1$ limit}

We have shown that all physically realistic cases satisfy~$S\ll1$, but we
nevertheless consider the limit of~$S=1$ (corresponding to~$\mu=M$) in order to
investigate more thoroughly the dynamics of the Papapetrou equations, and to
compare our results with~\cite{SuzukiMaeda1997}, which investigates the~$S=1$
limit in detail.  The~$S=1$ limit introduces no singularities into the
equations of motion, and the resulting orbits are valid solutions of the
equations.  On the other hand, in this limit the Papapetrou equations are not
physically  realistic, since they are derived in the limit of spinning test
particles, which must satisfy $\mu\ll M$. We thus cannot draw reliable results
about the behavior of astrophysical systems from the  $S=1$ limit.

\subsection{Symmetries and the parameterization of initial conditions}
\label{sec:parameterization}

In the approximation represented by the Papapetrou equations
there is still a constant
of the motion associated with each Killing vector~$\vec\xi$
of the spacetime~\citep{Dixon}:
\begin{equation}
C_{\xi}=\xi^\mu p_\mu-\textstyle{1\over2}\xi_{\mu;\nu}
S^{\mu\nu}.
\end{equation}
[For brevity, we write the constant in terms of the spin 
tensor~$S^{\mu\nu}$~[Eq.~(\ref{eq:S_tensor})].]
Since Kerr spacetime is stationary and axially symmetric,
it has the Killing vectors
$\vec\xi^{t}=\partial/\partial t$ and $\vec\xi^{\phi}=\partial/\partial\phi$, 
so the energy~$E$ and $z$~angular momentum~$J_z$ are conserved: 
\begin{equation}
E=-p_t+\textstyle{1\over2}g_{t\mu,\nu}S^{\mu\nu}
\end{equation}
and
\begin{equation}
J_z=p_\phi-\textstyle{1\over2}g_{\phi\mu,\nu}S^{\mu\nu}.
\end{equation}
(We write $J_z$ in place of the orbital angular momentum~$L_z$ since
the spin also contributes to the angular momentum of the system.)
In contrast to the energy and momentum integrals, the Carter constant~$Q$
is no longer present when the test particle has nonvanishing
spin~\citep{TMSS1996}. 

In our problem there are twelve variables, four each for position, momentum,
and spin.
For the purposes of finding orbits by numerical integration, we may parameterize
the initial conditions by providing $\tau=0$, $r$, $\theta$, $\phi=0$,
 $p_r$, $E$, $J_z$, $S$, and any two
of the spin components. The normalization conditions $p_\mu p^\mu=-1$ and
$S_\mu S^\mu=S^2$ allow us to eliminate one component each of momentum 
and spin.  The constraint~$p^\mu S_\mu=0$ and the
integrals of the motion then
give three equations in three unknowns:
\begin{eqnarray}
0&=&p_\mu S_\nu g^{\mu\nu}\\
E&=&-p_t+\textstyle{1\over2}g_{t\mu,\nu}S^{\mu\nu}\\
J_z&=&p_\phi-\textstyle{1\over2}g_{\phi\mu,\nu}S^{\mu\nu}
\end{eqnarray}
We must solve these equations for the two remaining components of $p_\mu$ and
one remaining
component of~$S_\mu$. 
In Schwarzschild spacetime
these can be solved explicitly due to the greater symmetry~\citep{SuzukiMaeda1997}, but
in the Kerr case of interest here
the problem requires numerical root finding.

We also use a related parameterization method starting with the Kerr geodesic
orbital parameters: eccentricity~$e$, inclination angle~$\iota$, and
pericenter~$r_p$. We derive the corresponding energy, angular momentum, and
relevant momenta, and then proceed as above. This method is discussed further
in Sec.~\ref{sec:kerr_schw}.

\section{Lyapunov exponents}

\subsection{General discussion of Lyapunov exponents}
\label{sec:lyapunov_exponents}

Our method for calculating Lyapunov exponents is well-established in the
literature of nonlinear dynamical systems~\citep{ASY1997,Ott1993}, but accessible
treatments are hard to find in the physics literature, so we summarize the method
here. Our discussion is informal and oriented toward practical calculation, based
on Ref.~\citep{ASY1997}; for 
a more formal, rigorous presentation see Eckmann and Ruelle~\citep{EckmannRuelle1985}.

First we give an overview of the methods for calculating Lyapunov exponents most commonly
used in physics. Given an initial condition, a set of differential equations determines
a solution (the \emph{flow}), which is a curve in the phase space.
The \emph{Lyapunov exponents} of the flow
measure the rate at which nearby trajectories separate. As discussed
in the introduction, an orbit is chaotic if a nearby phase-space trajectory separated
by an initial distance~$\epsilon_0$ 
separates exponentially: 
$\epsilon(\tau)=\epsilon_0\,e^{\lambda \tau}$, where $\lambda$ is the Lyapunov exponent.

Implicit in the definition of chaos
above is a notion of a distance function on the phase space (or, more
properly, the tangent space to the phase space, as in~Eq.~(\ref{eq:dxidt}) below).
It is conventional to use a Euclidean metric to define such lengths~\cite{ASY1997,Ott1993},
but any positive-definite nondegenerate metric will 
do~\citep{EckmannRuelle1985}. While the magnitude of the resulting
exponent obviously depends on the particular
metric used, the signs of the Lyapunov exponents
are a property of the dynamical system and do not rely on any underlying metric 
structure. We discuss these issues further in Sec.~\ref{sec:projnorm}
and Sec.~\ref{sec:comments}.

This informal definition of Lyapunov exponents leads to a practical method for calculating
$\lambda$: given an initial condition, consider a nearby initial
condition a distance $\epsilon_0$ away, where $\epsilon_0$ is ``small'', typically 
$10^{-5}$--$10^{-7}$ of the relevant physical scales. (Values of $\epsilon_0$ 
much smaller than this can result in a loss of numerical precision.)
Keeping track of the deviation vector between the two
points yields a numerical
approximation of~$\lambda$. (It is important to rescale the deviation vector if 
it grows too large, since for any bounded phase space flow
even a tiny deviation can grow to at most the size of the bounded region.)
We call this approach the \emph{deviation vector method}.

There are two primary limitations to the approach outlined above. First, the
method yields only the largest Lyapunov exponent, which is sufficient to
establish the presence  of chaos but paints a limited picture of the dynamics.
Second, the deviation vector approach is most appropriate when an analytical
expression for  the Jacobian matrix is unknown; by choosing $\epsilon_0$ small
enough (and by keeping $\epsilon(\tau)$ small by rescaling if necessary), the
method essentially takes a numerical derivative. Among other complications, the
value of the exponent depends both on the maximum allowable
size~$\epsilon_\mathrm{max}$ (the size at which the deviation is rescaled) and
the initial value $\epsilon_0$ (the size of the deviation after each
rescaling). 

The principal virtue of the deviation vector approach compared to the more
complicated Jacobian method (discussed below) is speed, since  it
requires solving only the equations of motion. (As we discuss in
Sec.~\ref{sec:algorithm_detail}, the Jacobian method involves the
time-consuming evolution of the Jacobian matrix in parallel with the equations
of motion.) It also provides a valuable way to verify the validity of the
Jacobian method.

The \emph{Jacobian method} is a more thorough and rigorous approach to the
calculation of Lyapunov exponents, which  makes precise the notion of 
``infinitesimally'' separated vectors.  The general method proceeds
as follows: consider a
phase space with variables ${\bf y}=\{y_i\}$ and an autonomous set of
differential equations 
\begin{equation}
\label{eq:f}
\frac{d\mathbf{y}}{d\tau}=\mathbf{f}(\mathbf{y}).
\end{equation}
(Here we use~$\tau$ instead of~$t$ in anticipation of the application of these results to
general relativity, where we will be using proper time as our time parameter.)
If $\delta{\bf y}$ represents a small deviation vector, then 
the distance between the two trajectories is 
\begin{equation}
\frac{d(\delta{\bf y})}{d\tau}={\bf f}({\bf y}+\delta{\bf y})-{\bf f}({\bf y})={\bf Df}\cdot
\delta{\bf y}+{\cal O}(\|\delta{\bf y}\|^2),
\end{equation}
where ${\bf Df}$ is the Jacobian matrix 
[$({\bf Df})_{ij}=\partial f_i/\partial x^j$]. 

We can clarify the notation and make the
system easier to visualize if we introduce
$\bm{\xi}$ as an element of the tangent space at ${\bf y}$, so that 
\begin{equation}
\label{eq:dxidt}
\frac{d\bm{\xi}}{d\tau}={\bf Df}\cdot\bm{\xi},
\end{equation}
which is equivalent to taking the limit $\|\delta{\bf y}\|\rightarrow0$. We
visualize $\bm{\xi}$ as a perfectly finite vector (as opposed to an 
``infinitesimal'').
Since it lives in the tangent space, not the physical phase space, $\bm{\xi}$
can grow arbitrarily large with time. This means that instead 
of the frequent rescaling required in the deviation vector approach, 
$\bm{\xi}$ must be rescaled only 
when it grows so large that it approaches the floating point limit of the computer.
This is a rare occurrence, and in practice the tangent vector almost never needs rescaling.

\begin{figure}
\includegraphics[width=3.in]{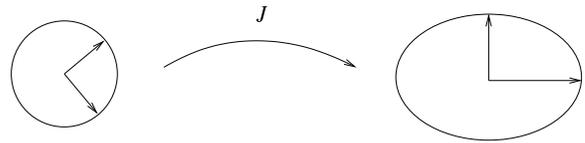}
\caption{\label{fig:jacobian}
The Jacobian matrix maps a tangent space ball to an ellipsoid.}
\end{figure}

Although following the evolution of an arbitrary initial tangent vector
$\bm{\xi}$  yields the largest Lyapunov exponent, we can do even better by
following the evolution of a family of~$n$ tangent vectors, which allows us to
determine all~$n$ Lyapunov exponents. The essence of the method is as follows:
for a system of differential equations with  $n$~variables, we consider a set
of $n$~vectors that lie on a ball in the tangent space. We represent this ball
using a matrix whose columns are $n$~normalized, linearly independent tangent
vectors, conventionally taken to be orthogonal.  This set of orthonormal
vectors then spans a unit ball in the tangent space. The action of the Jacobian
matrix, which is a linear operator on the tangent space, is to map the ball to
an ellipsoid under the time-evolution of the flow, as shown in
Fig.~\ref{fig:jacobian}.

For a dynamical system with $n$~degrees of freedom, there are~$n$ 
\emph{Lyapunov numbers} that measure the average growth of the $n$~principal 
axes $\{r_i(\tau)\}_{i=1}^n$ of the ellipsoid. 
More formally, the Lyapunov numbers $L_i$ are
given by
\begin{equation}
L_i=\lim_{\tau\rightarrow\infty} [r_i(\tau)]^{1/\tau},
\end{equation}
where $r_i(\tau)$ is the length of the $i$th principal axis of the ellipsoid. 
The corresponding
Lyapunov exponents are the natural logarithms of the Lyapunov
numbers, so that 
\begin{equation}
\lambda_i=\lim_{\tau\rightarrow\infty}\frac{\log\,[r_i(\tau)]}{\tau}.
\end{equation}
These limits exist for a broad class of dynamical systems~\citep{EckmannRuelle1985}.

The principal axes of the tangent space 
ellipsoid indicate the directions along which nearby initial
conditions separate or converge, which we may call the \emph{Lyapunov directions}. 
In particular, consider a principal axis that is
stretched under the time evolution.  Such a vector has one component for each dimension
(position or momentum) in the phase space; a nonzero component in any direction indicates
an exponential divergence in the corresponding coordinate. For example, if a system
has two spatial coordinates $(r, \phi)$ and corresponding momenta $(p_r, p_\phi)$, then
a typical tangent vector will have components
$\bm{\xi}=(\xi_r,\xi_\phi,\xi_{p_r},\xi_{p_\phi})$.  If the only tangent vector
with nonzero Lyapunov exponent is, for example,
$\bm{\xi}={1\over\sqrt{3}}(1,1,1,0)$, then nearby initial conditions
separate exponentially in $r$, $\phi$, and $p_r$, but nearby values of
$p_\phi$ do not separate exponentially.  This is potentially relevant
to the present study since, in the limit of a point test particle, the
gravitational radiation depends on the spatial variables but not the spin.
If the principal axes along expanding directions have nonzero components
only in the spin directions, the system could be formally chaotic without
affecting the gravitational waves.

In summary, the method for visualizing the Lyapunov exponents of
a dynamical system is to picture a ball of initial conditions---an
infinitesimal ball if visualized in the phase space, or a unit ball if
visualized in the tangent space---and watch it evolve into an ellipsoid
under the action of the Jacobian matrix. After a sufficiently long
time, the ellipsoid will be greatly deformed, stretched out along the
expanding directions and compressed along the contracting directions. The
directions of the principal axes are the Lyapunov directions, and their
lengths give the Lyapunov numbers through the relation $L_i\approx
[r_i(\tau)]^{1/\tau}$.

\subsection{Numerical calculation of Lyapunov exponents}
\label{sec:num_lyap}

In order to implement a numerical algorithm based on the considerations above,
we must bear two things in mind. First, since the vectors spanning the initial
unit ball are arbitrary, they will all be stretched in the direction of the
largest exponent: in general every initial vector has some nonzero
component along the direction of greatest stretching, which dominates as 
$\tau\rightarrow\infty$. In order to find the other principal axes, we must
periodically produce a new orthogonal basis. We will show that the Gram-Schmidt
procedure is appropriate. Second, the lengths of the vectors
could potentially overflow or underflow the machine precision, so we should
periodically normalize the ellipsoid axes. 

\subsubsection{The algorithm in detail}
\label{sec:algorithm_detail}

To simplify the notation, we denote the (time-dependent) Jacobian matrix $\mathbf{Df}$
by $J_\tau$ and the ellipsoid (whose columns are the tangent vectors) by $U$.
The algorithm then proceeds as follows:
\begin{enumerate}

\item Construct a set of $n$~orthonormal vectors (which span an
$n$-dimensional ball in the tangent space of the flow). Represent
this ball by a matrix $U$ whose columns are the tangent vectors
$\bm{\xi}_i$.

\item Eq.~(\ref{eq:dxidt}), applied to each tangent vector, implies that $U$
satisfies the matrix equation
\begin{equation}
\label{eq:dUdt}
\frac{dU}{d\tau}=J_\tau\,U,
\end{equation}
which constitutes a set of \emph{linear} differential 
equations for the tangent vectors.
Since $J_\tau$
depends on the values of $\mathbf{y}$,
these equations are coupled to our system of nonlinear differential equations 
$\dot{\mathbf{y}}=\mathbf{f}(\mathbf{y})$, so they must be solved in parallel with 
Eq.~(\ref{eq:f}).

\item Choose some time~$T$ big enough to allow the expanding directions to grow
but small enough so that they are not too big. 
Numerically integrate Eqs.~(\ref{eq:f}) and~(\ref{eq:dUdt}), and every time~$T$ apply
the Gram-Schmidt orthogonalization procedure.
The vectors resulting from the Gram-Schmidt procedure approximate
the semiaxes of the evolving ellipsoid. Record the log of the length 
$\log\,[r_i(\tau_n)]$ 
of each vector after each time~$T$, where~$\tau_n=nT$.
Finally, normalize the ellipsoid back to a unit ball.

\item At each time~$\tau$, the sum
\begin{equation}
\label{eq:log_r}
\lambda_i\approx{1\over \tau}\sum_{n=1}^{N}\log\,[r_i(\tau_n)]
\equiv\frac{\log\,[r_i(\tau)]}{\tau}
\end{equation}
is a numerical estimate for the $i$th~Lyapunov exponent.  
\end{enumerate}

\subsubsection{Gram-Schmidt and Lyapunov exponents}

The use of the Gram-Schmidt procedure is crucial to extracting all $n$ Lyapunov
exponents. Let us briefly review this important construction. Given $n$ 
linearly-independent vectors $\{{\bf u}_i\}$, the Gram-Schmidt procedure constructs
$n$ orthogonal vectors $\{{\bf v}_i\}$ that span the same space, given by
\begin{equation}
{\bf v}_i={\bf u}_i-\sum_{j=1}^{i-1}\frac{{\bf u}_i
\cdot{\bf v}_j}{\|{\bf v}_j\|^2}\,{\bf v}_j.
\end{equation}
To construct the $i$th orthogonal vector, we take the $i$th vector from the original set
and subtract off its projections onto the previous $i-1$ vectors produced by
the procedure.

The use of Gram-Schmidt in dynamics comes from observing that the resulting
vectors approximate the semiaxes of the tangent space ellipsoid. 
After the first time~$T$, all of the vectors point mostly along the
principal expanding direction.  We may therefore pick any one as
the first vector in the Gram-Schmidt algorithm, so choose
$\bm{\xi}_1\equiv{\bf u}_1$ without loss of generality. If we let ${\bf e}_i$ denote
unit vectors along the principal axes and let $r_i$ be the lengths of
those axes, the dynamics of the system
guarantees that the first vector ${\bf u}_1$ satisfies
\[
{\bf u}_1=r_1{\bf e}_1+r_2{\bf e}_2+
\cdots\approx r_1{\bf e}_1\equiv{\bf v}_1
\]
since ${\bf e}_1$ is the direction of fastest stretching. The second
vector ${\bf v}_2$ given by Gram-Schmidt is then 
\[
{\bf v}_2={\bf u}_1-\frac{{\bf u}_1\cdot{\bf v}_1}{\|{\bf v}_1\|^2}\,{\bf v}_1
\approx{\bf u}_1-r_1{\bf e}_1=r_2{\bf e}_2,
\]
with an error of order $r_2/r_1$. The procedure proceeds iteratively, with
each successive Gram-Schmidt step (approximately) subtracting off the contribution
due to the previous semiaxis direction.

It is important to choose time~$T$ long enough to keep errors of the
form~$r_2/r_1$ small but short enough to prevent numerical under- or overflow. 
In practice, the method is quite robust, and it is easy to find valid choices
for the time~$T$, as discussed in Sec.~\ref{sec:results}.

\section{Relativity and Papapetrou subtleties}

The algorithm described above is of a general nature, designed with
a generic dynamical system in mind.  The Papapetrou equations and the
framework of general relativity present additional complications.
Here we discuss some refinements to the algorithm necessary for the present
case.

\subsection{Phase space norm}
\label{sec:projnorm}

In the context of general relativistic dynamical systems, the meaning of trajectory
separation in phase space is somewhat obscured by the time variable.  
We can skirt the issue of trajectories ``diverging in time'' by using
a $3+1$ splitting of spacetime, and consider trajectory separation in a spacelike
hypersurface~\citep{KarasVokrouhlicky1992}.  This prescription reduces properly to the 
traditional method for classical dynamical systems in the nonrelativistic limit.

In Kerr spacetime, we use the zero angular-momentum observers (ZAMOs), and
project 4-dimensional quantities into the ZAMO hypersurface using the
projection tensor $P^\mu_{\ \nu} = \delta^\mu_{\ \nu} + U^\mu U_\nu$, where
$U^\mu$ is the ZAMO 4-velocity. In this formulation,  spatial variables obey
$x^\mu \rightarrow \tilde x^i = P^i_{\ \mu}\,x^\mu$ and momenta obey $p_\mu
\rightarrow \tilde p_i = P^\mu_{\ i}\,p_\mu$ (and similarly for
$S_\mu$)~\citep{KarasVokrouhlicky1992}.   The relevant norm is then a Euclidean
distance in the 3-dimensional hypersurface.

We should note that we use the projected norm for conceptual clarity, not necessity.
The naive use of a Euclidean norm using unprojected components yields
the same sign for the exponents, as noted in 
Sec.~\ref{sec:lyapunov_exponents}.  The magnitudes of the resulting exponents are also
similar~(Sec.~\ref{sec:comments}).

\subsection{Constraint complications}
\label{sec:constraint_complexities}

Although the Lyapunov algorithm is fairly straightforward to implement for
a general dynamical system, the constrained nature of the Papapetrou equations
adds a considerable amount of complexity. The fundamental problem is that the 
tangent vector~$\bm{\xi}$ cannot have arbitrary initial components for the Papapetrou
system, as it can for an unconstrained dynamical system.
Each~$\bm{\xi}$ must correspond to some 
deviation~$\delta\mathbf{y}$ which is not arbitrary: the deviated point
$\mathbf{y} + \delta\mathbf{y}$ must satisfy the constraints.

\subsubsection{Constraint-satisfying deviations}
\label{sec:constraints}

Recall that the dynamical variables in the Papapetrou equations must satisfy
normalization and orthogonality constraints (Sec.~\ref{sec:Papapetrou_eq}): 
$p^\nu p_\nu = -1$ (normalized units), $S^\nu S_\nu = S^2$, and $p^\nu S_\nu = 0$.
To make the constraint condition on~$\delta{\bf y}$
clearer, let $\mathbf{C}\mathbf(\mathbf{y})=0$ represent the constraints
rearranged so that the right hand side is zero.  For example, with ${\mathbf{y}} = (t, r,
\mu, \phi, p_t, p_r, p_\mu, p_\phi, S_t, S_r, S_\mu, S_\phi)$,\footnote{Recall
that we write the equations of motion in terms of $\mu=\cos\theta$.}
we can write 
\begin{equation}
C_1({\mathbf{y}}) = p^\nu p_\nu + 1,
\end{equation}
so that $C_1({\mathbf{y}}) = 0$ for a constraint-satisfying ${\mathbf{y}}$.  
The other constraints are then
\begin{equation}
C_2({\mathbf{y}}) = S^\nu S_\nu - S^2
\end{equation} and 
\begin{equation}
C_3({\mathbf{y}}) = p^\nu S_\nu.
\end{equation}
A deviation $\delta\mathbf{y}$ is \emph{constraint-satisfying} if 
${\mathbf{C}}({\mathbf{y}}+\delta{\mathbf{y}})={\bf 0}$ when 
${\mathbf{C}}({\mathbf{y}})={\bf 0}$.

We may construct a constraint-satisfying deviation~$\delta{\mathbf{y}}$
as follows.  Begin with a 12-dimensional vector ${\bf y}$ that satisfies
the constraints.  Add a random small deviation to eight of its components to
form a new vector ${\bf y}'$. (We need not add a deviation to~$t$; see
Sec.~\ref{sec:modified_GS} below.) Determine the remaining three components of
${\bf y}'$ using the constraints, using the same technique used
to set the initial conditions.  Finally, set $\delta{\bf y} \equiv
{\bf y}' - {\bf y}$.  The corresponding~$\bm{\xi}$ is then simply
$\delta{\mathbf{y}}/\|\delta{\mathbf{y}}\|$.

The prescription above glosses over an important detail: the inference of tangent vector
components from the constraints is not unique.  Solving the constraint equations involves
taking square roots in several places, so there are a number of sign ambiguities representing
different solution branches. 
The implementation of the
component-inference algorithm must compare each component of ${\bf y}$ with the corresponding
component of ${\bf y}'$ to ensure that they represent solutions from the same branches.
Enforcing the constraints in this manner, 
and thereby inferring the full tangent vector~$\bm{\xi}$, is
especially important for the algorithm described in the next section.

\subsubsection{A modified Gram-Schmidt algorithm}
\label{sec:modified_GS}

A spinning test particle has an apparent twelve
degrees of freedom---four each for position, momentum, and spin---so \emph{a
priori} there is the potential for twelve nonzero exponents. 
Since the Papapetrou equations have no explicit time-dependence, 
we can eliminate the time degree of freedom.
The three constraints (momentum and spin normalization,
and momentum-spin orthogonality) further reduce the number of degrees of freedom by
three.  We are left finally with eight degrees of freedom.

Eliminating the four spurious degrees of  freedom from the tangent vectors
presents a formidable obstacle to the implementation of the phase space
ellipsoid method described in Sec.~\ref{sec:algorithm_detail}. The crux of the
dilemma is that the axes of the ellipsoid must be orthogonal, but must also
correspond to constraint-satisfying deviation vectors---mutually  exclusive
conditions. Solving this problem requires a modification of the Gram-Schmidt
algorithm:

\begin{enumerate}

	\item Instead of a $12\times12$ ball (i.e., $n=12$ in the original algorithm),
	consider an $8\times8$ ball by choosing to eliminate the $t$,
	$p_t$, $p_\phi$, and $S_t$ components.  The time component~$\xi_t$ of each
	tangent vector is irrelevant since
	nothing in the problem is explicitly time dependent; the first column of the
	Jacobian matrix is zero, so~$\xi_t$ is not necessary to determine the
	time-evolution.\footnote{Also, the time piece is discarded in the projected norm
	formalism in any case (Sec.~\ref{sec:projnorm}).}  The other three
	components are determined by the constraints as described above.

	\item Given eight initial random tangent vectors, apply the
	Gram-Schmidt process to form an
	$8\times8$~ball. For each vector, determine the three missing components using
	the constraints, and then evolve the system using \[ \frac{dU}{d\tau} =
	J_\tau U\] as before. (Now $U$ represents a $12\times8$ matrix instead of
	a $12\times12$~ball.)

	\item At each time~$T$, extract the relevant eight components from each
	vector to form a new $8\times8$~ellipsoid, apply Gram-Schmidt, and then
	fill in the missing components using the constraints, yielding again a
	$12\times8$~matrix.  The projected norms of the eight tangent vectors
	contribute to the running sums for the Lyapunov exponents as in the
	original algorithm.

\end{enumerate}
The algorithm above yields eight Lyapunov exponents for the Papapetrou system of equations.

In order to implement this algorithm, we must have a method for constructing a full
tangent vector~$\bm{\xi}$ from an eight-component vector~$\tilde{\bm{\xi}}$. The
method is as follows:
\begin{enumerate}

	\item Let $\tilde{\bf y}' = {\bf y} + \epsilon\tilde{\bm{\xi}}$ for a suitable choice
	of~$\epsilon$.

	\item Fill in the missing components of~$\tilde{\bf y}'$ using the constraints to
	form~${\bf y}'$, taking care that ${\bf y}$~and~${\bf y}'$ have the same
	constraint branches.

	\item Infer the full tangent vector using $\bm{\xi} =
	\displaystyle{\frac{{\bf y}' - {\bf y}}{\epsilon}}$.

\end{enumerate}
This technique depends on the choice of~$\epsilon$, and fails when~$\epsilon$
is too small or too large.  Using the techniques
discussed in the next section to calibrate the system, we find that 
$\epsilon\approx10^{-5}$--$10^{-6}$ works well in practice.

\subsubsection{Two rigorous techniques}

It should be clear from the discussion above that extracting all eight Lyapunov exponents is
difficult, and in practice the techniques are finicky, depending (among other
things) on the choice of~$\epsilon$ as described in Sec.~\ref{sec:modified_GS} above. 
How, then, can we be confident that the results make sense?  Fortunately,
there are two techniques that give rigorous Lyapunov exponents by managing
to sidestep the constraint complexities entirely.

First, it is always possible to calculate the single largest exponent
using the Jacobian method without considering the constraint subtleties.  
The complexity of the
main Jacobian approach involves the competing requirements of Gram-Schmidt
orthogonality and constraint satisfaction, but in the case of only
one vector these difficulties vanish.  Since the equations of motion
preserve the constraints, an initial constraint-satisfying tangent
vector retains this property throughout the integration.  Thus, we begin
with a vector constructed as in Sec.~\ref{sec:constraints} and evolve
it (without rescaling) along with the equations of motion. 
Other
than the requirement of constraint satisfaction, its initial components are arbitrary, so
it evolves in the direction of largest stretching and eventually points
in the largest Lyapunov direction.  The logarithm of its projected norm
then contributes to the sum for the largest Lyapunov exponent.

Second, we can implement a deviation vector approach as described in
Sec.~\ref{sec:lyapunov_exponents}.  Given an initial condition~${\bf y}_0$, we construct a nearby
initial condition~${\bf y}'_0$ as in Sec.~\ref{sec:constraints} and then evolve them both
forward. In principle, an approximation for the largest Lyapunov exponent is then
$\displaystyle
{
	{1\over\tau}\log
	\left(
		\frac
		{
			\| {\bf y}'-{\bf y} \|
		}
		{
			\| {\bf y}'_0 - {\bf y}_0 \|
		}
	\right)
	\equiv 
	{1\over\tau}\log
	\left(
		\frac{\|\delta{\bf y}\|}{\|\delta{\bf y}_0\|}
	\right)
}$.  In practice (for chaotic systems) the method saturates:
for a given initial
deviation, say $\|\delta{\bf y}_0\|\sim10^{-6}$, once the initial conditions have diverged by a
factor of~$\sim\!\!10^6$ the method breaks down.\footnote{This underscores the point that
chaos is essentially a \emph{local} phenomenon.  \emph{Any} unrescaled deviation 
vector approach must saturate, since no bounded system can have trajectories that diverge
for arbitrarily long times.}
(The traditional solution to the
saturation problem
is to rescale the deviation before it saturates, 
but such a rescaling in this case violates the constraints.)  
Despite its limitations, this unrescaled
deviation vector technique is valuable, since it tracks the correct solution until
the saturation limit is reached, and avoids the subtleties associated with the constraints. 

With these two techniques in hand, we have a powerful method for verifying that the
largest Lyapunov exponent produced by the Gram-Schmidt method is correct.  This, in turn,
gives us confidence that the other Lyapunov exponents produced by the main algorithm are
meaningful as well.

\section{Implementation details}

\subsection{Some numerical comments}
\label{sec:numerical}

Finally, we discuss some specialized
issues related to integrating the Papapetrou equations on
a computer.  The primary subjects are the formulation of the equations, optimization
techniques,
and error checking.

Our choice to write the Papapetrou equations using the spin vector is 
motivated partially by numerical considerations. The spin vector approach
has nice properties compared to the tensor approach as $S\rightarrow0$.
Comparing their covariant derivatives is instructive:
\begin{eqnarray*}
\nabla_{\vec v}\,S_\mu&=&-p_\mu \left(R^{*\alpha\ \gamma\delta}_{\ \ \beta}
	S_\alpha v^\beta p_\gamma S_\delta\right)\\
\nabla_{\vec v}\,S^{\mu\nu}&=&p^\mu v^\nu-p^\nu v^\mu = 2p^{[\mu}v^{\nu]}.
\end{eqnarray*}
Though simpler in form, the derivative of $S^{\mu\nu}$ has unfortunate numerical
properties for small~$S$, since in the limit $S\rightarrow0$ we have 
$p^\mu\rightarrow v^\mu$:
the difference $p^\mu v^\nu-p^\nu v^\mu$ goes to zero in principle but in practice is
plagued by numerical roundoff errors. 
Since $S\ll1$ is the most physically
interesting limit, the vector approach is more convenient for our purposes.

Calculating the 
many tensors and derivatives which go into the Papapetrou equations
and the corresponding Jacobian matrix is a considerable task.
As a first step, we use GRTensor for Maple to calculate
all relevant quantities, and we use Maple's optimized~C output to
create C~code automatically.  
Due to the symmetries of the Riemann tensor and
the metric, many terms are identically zero, which significantly reduces
the number of required operations.  For example, in order to
calculate $R^{*\alpha\ \gamma\delta}_{\ \ \beta}
	S_\alpha v^\beta p_\gamma S_\delta$ we need four loops, which
	constitutes
$4^4=256$ evaluations, but in fact $R^{*\alpha\ \gamma\delta}_{\ \ \beta}$
has only 80 nonzero components. Performing loop unrolling by 
writing these terms to an optimized
derivatives file consisting of explicit sums 
speeds up calculation by an order of magnitude
compared to nested {\tt for} loops.

Another optimization involves
the choice of coordinates used in the metric, which has significant consequences
for the size of the tensor files and the number of floating point operations required.
Simply using $\mu=\cos\theta$ in the Kerr metric reduces the size of the Riemann 
derivatives by at least a factor of two.\footnote{{\sl Warning:} This variable
substitution changes the handedness of the coordinate system, since the unit 
vector~$\hat\mu$ points opposite to~$\hat\theta$.  This in turn introduces an extra minus
sign in the Levi-Civita tensor~$\epsilon^{\alpha\beta\gamma\delta}$, which appears many
times in the Papapetrou equations and the corresponding conserved quantities.  The author
discovered this subtlety the hard way.}
Since these derivatives are the bottleneck
in the calculation of the Jacobian matrix, we can get more than a 50\% improvement
in performance with even this simple variable transformation.

All integrations were performed using a Bulirsch-Stoer integrator adapted
from \emph{Numerical Recipes}~\citep{NumRec}. Occasional checks with a
fifth-order Runge-Kutta integrator were in agreement. We verified the Papapetrou
integration by checking errors in the constraints and conserved quantities; for
an orbit such as that shown in Fig.~\ref{fig:kerr_spin_orbit}, all
errors are at the $10^{-11}$ level after $\tau = 10^5\,M$.

As should be clear from Sec.~\ref{sec:Jacobian_matrix} below, the  Jacobian
matrix of the Papapetrou equations has a large number of terms, and it is
essential to verify its correctness by using a diagnostic that compares
$\mathbf{Df}\cdot\delta\mathbf{y}$ with the difference
$\mathbf{f}(\mathbf{y}+\delta\mathbf{y}) -\mathbf{f}(\mathbf{y})$ for a
suitable constraint-satisfying $\delta\mathbf{y}$. It is not sufficient for the
difference merely to be small:   we must calculate the quantity 
$\mathbf{f}(\mathbf{y}+\delta\mathbf{y}) -\mathbf{f}(\mathbf{y}) - {\bf
Df}\cdot\delta{\bf y}$ for  several values of $\delta{\bf y}$ and verify that
each component scales as~$\|\delta{\bf y}\|^2$.   An early implementation of
the Jacobian matrix, which gave nearly identical results for
$\mathbf{f}(\mathbf{y}+\delta\mathbf{y}) -\mathbf{f}(\mathbf{y})$ and ${\bf
Df}\cdot\delta{\bf y}$, nevertheless had an undetected~${\cal O}(S^2)$ error. 
The unrescaled deviation vector approach showed a discrepancy with the Jacobian
method,\footnote{This illustrates the value of calculating the Lyapunov
exponents using two different methods.} which showed spurious chaotic behavior.
The $\|\delta{\bf y}\|^2$~scaling method described above  eventually diagnosed
the problem, which resulted from a missing term in $\partial\dot S_\mu/\partial
S_\nu$ (Sec~\ref{sec:Jacobian_matrix}).

\subsection{The Jacobian matrix}
\label{sec:Jacobian_matrix}

For reference, we write out explicit equations for part of the Jacobian matrix of
the Papapetrou equations. 

The Jacobian matrix of a system of differential equations, specialized to the case at hand, is as follows: 
\begin{equation}
\label{eq:jacobian}
\left(
\begin{array}{rcl}
	\displaystyle{\frac{\partial\dot x^\mu}{\partial x^\nu}} & 
		\displaystyle{\frac{\partial\dot x^\mu}{\partial p_\nu}} &
		\displaystyle{\frac{\partial\dot x^\mu}{\partial S_\nu}}\medskip\\
	\displaystyle{\frac{\partial\dot p_\mu}{\partial x^\nu}} & 
		\displaystyle{\frac{\partial\dot p_\mu}{\partial p_\nu}} &
		\displaystyle{\frac{\partial\dot p_\mu}{\partial S_\nu}}\medskip\\
	\displaystyle{\frac{\partial\dot S_\mu}{\partial x^\nu}} & 
		\displaystyle{\frac{\partial\dot S_\mu}{\partial p_\nu}} &
		\displaystyle{\frac{\partial\dot S_\mu}{\partial S_\nu}}\\
\end{array}
\right)
\end{equation}
Once we calculate $\partial\dot x^\mu/\partial x^\nu=v^\mu_{\ ,\nu}$, all the other
derivatives can be expressed in terms of the derivatives of $v^\mu$,
the tensors and connection coefficients, and Kronecker $\delta$s.

Written out in full, and the Papapetrou equations are as follows:
\begin{eqnarray}
	\dot x^\mu & = & v^\mu\\
	\dot p_\mu & = & -R^{*\ \ \alpha\beta}_{\mu\nu}v^\nu p_\alpha S_\beta+
		\Gamma^\alpha_{\ \beta\mu}p_\alpha v^\beta\\
	\dot S_\mu & = & -p_\mu \left(R^{*\alpha\ \gamma\delta}_{\ \ \beta} S_\alpha v^\beta p_\gamma S_\delta\right)+
		\Gamma^\alpha_{\ \beta\mu}S_\alpha v^\beta
\end{eqnarray}
We measure $\tau$ and $r$ in units of~$M$ (the mass of the central body), $p_\mu$
in units of the particle rest mass~$\mu$, and $S_\mu$ in terms of the product $\mu M$.
The overdot is an ordinary derivative with respect to proper time: $\dot x\equiv
dx/d\tau$.

The unusual placement of indices on $R^*$ is motivated by the form of the Jacobian
matrix.  The index placement shown above
brings the equations into a form where the indices on $p_\mu$ and $S_\mu$ are
always lowered, which simplifies the Jacobian matrix since (for example) $\partial p_\mu/
\partial x^\mu=0$.
Otherwise the Jacobian matrix is unnecessarily complicated; for example,
if $p^\mu$ appeared anywhere on the right hand side then we would have
$\partial p^\mu/\partial x^\nu\neq0$, which would contribute to~$J_\tau$.

As discussed in Sec~\ref{sec:Papapetrou_eq}, 
the supplementary condition
$p_\mu S^{\mu\nu}=0$ [Eq.~(\ref{eq:p_ssc})] leads 
to the equation for $v^\mu$ in terms of $p^\mu$:
\begin{equation}
	v^\mu=N(p^\mu+w^\mu) = 
		N\tilde v^\mu,
\end{equation}
where
\begin{equation}
\tilde v^\mu = p^\mu+w^\mu
\end{equation}
and
\begin{equation}
w^\mu=-{^*}R^{*\mu\alpha\beta\gamma} S_\alpha p_\beta S_\gamma.
\end{equation}
$N$~is a normalization constant fixed by $v_\mu v^\mu = -1$. 

The calculation of the partial derivatives $\dot x^\mu$ in 
Eq.~(\ref{eq:jacobian})
proceeds as follows.
From the relation for $v^\mu = N\tilde v^\mu$, we have 
\[\frac{\partial\dot x^\mu}{\partial x^\nu} = 
v^\mu_{\ ,\nu}=N\tilde v^\mu_{\ ,\nu}+N_{,\nu}\tilde v^\mu.\] Now, $\tilde v^\mu_{\ ,\nu}
=p^\mu_{\ ,\nu}+w^\mu_{\ ,\nu}=p_\alpha g^{\alpha\mu}_{\ \ ,\nu}
-{^*}R^{*\mu\alpha\beta\gamma}_{\ \ \ \ \ \ ,\nu}S_\alpha p_\beta S_\gamma$, so the 
first term is easy.
The second term is trickier: from the expression for $v^\mu$, we have that
$-1=v^\mu v_\mu=N^2(p^\mu p_\mu + 2 w^\mu p_\mu + w^\mu w_\mu)=N^2(-1+2 w^\mu p_\mu + w^\mu w_\mu)$, so we
have
\[
N=(1-2 w^\mu p_\mu - w^\mu w_\mu)^{-1/2}.
\]
Differentiating gives
\[
\begin{array}{rcl}
N_{,\nu}&=&N^3\left(p_\alpha w^\alpha_{\ ,\nu} + w^\alpha_{\ ,\nu} w_\alpha
	+\textstyle{1\over2}w^\alpha w^\beta g_{\alpha\beta,\nu}\right)\\
&	=&N^3\left(\tilde v_\alpha w^\alpha_{\ ,\nu}
	+\textstyle{1\over2}w^\alpha w^\beta g_{\alpha\beta,\nu}\right)
\end{array}\]
where we have re-labeled the dummy index ($\mu\rightarrow\alpha$).
Summing the various terms, we have
\begin{eqnarray}
v^\mu_{\ ,\nu}&=&N \left[p_\alpha g^{\alpha\mu}_{\ \ ,\nu} + w^\mu_{\ ,\nu} 
\nonumber\right.\\
&&\left. + \, v^\mu(v_\alpha w^\alpha_{\ ,\nu} +
\textstyle{1\over2}N w^\alpha w^\beta g_{\alpha\beta,\nu})\right].
\end{eqnarray}

The expression for $\partial\dot x^\mu/\partial p_\nu$ is similar to $v^\mu_{\ ,\nu}$, but it is 
simpler because the derivative of the metric with respect to the momentum is zero. As before, we use 
the product rule:
\[ \frac{\partial v^\mu}{\partial p_\nu}=N\frac{\partial\tilde v^\mu}{\partial p_\nu}
	+\frac{\partial N}{\partial p_\nu}\tilde v^\mu.
\] The first term requires
\[
\frac{\partial\tilde v^\mu}{\partial p_\nu}=\frac{\partial p^\mu}{\partial p_\nu}+
\frac{\partial w^\mu}{\partial p_\nu}=g^{\mu\nu}-
{^*}R^{*\mu\alpha\nu\beta}_{\ \ \ \ \ }S_\alpha S_\beta\equiv g^{\mu\nu}+W^{\mu\nu}.
\] Note that $W^{\mu\nu}$ is symmetric.
The second term requires
\[
\frac{\partial N}{\partial p_\nu}=N^3(W^{\alpha\nu}p_\alpha + w^\alpha\delta_\alpha^{\ \nu}
	+W^{\alpha\nu}w_\alpha)=N^3(w^\nu+\tilde v_\alpha W^{\alpha\nu}).
\] Summing the terms gives 
\begin{equation}
\frac{\partial \dot x^\mu}{\partial p_\nu}=
N(g^{\mu\nu}+W^{\mu\nu}+Nv^\mu w^\nu)+Nv^\mu v_\alpha W^{\alpha\nu}
\end{equation}
with
\begin{equation}
\frac{\partial w^\mu}{\partial p_\nu}\equiv W^{\mu\nu}=
-{^*}R^{*\mu\alpha\nu\beta}_{\ \ \ \ \ }S_\alpha S_\beta.
\end{equation}

Finally, we calculate $\partial\dot x^\mu/\partial S_\nu$.
With
\[
\frac{\partial\tilde v^\mu}{\partial S_\nu}=-S_\alpha p_\beta(
	{^*}R^{*\mu\alpha\beta\nu} - {^*}R^{*\mu\nu\alpha\beta})\equiv V^{\mu\nu},
\]
and 
\[
\frac{\partial N}{\partial S_\nu}\tilde v^\mu=Nv^\mu v_\alpha V^{\alpha\nu},
\] we have
\begin{equation}
\frac{\partial \dot x^\mu}{\partial S_\nu}=
NV^{\mu\nu}+Nv^\mu v_\alpha V^{\alpha\nu}.
\end{equation}

We calculate the derivatives of $\dot p_\mu$ and $\dot S_\mu$
using~$v^\mu_{\ ,\nu}$, the product rule, and the
derivatives of the various tensors in the problem. The full results appear
in the appendix.

\section{Integrability and chaos}
\label{sec:integrability}
\subsection{Phase space and constants of the motion}

Having laid the foundation for the numerical calculation of Lyapunov exponents,
we now discuss some general aspects of dynamical systems relevant to our study.
A dynamical system with $n$~coordinates has a $2n$~dimensional phase space, typically
consisting of generalized positions and their corresponding conjugate momenta. Motion
in the phase space is arbitrary in general, but when there are integrals of the 
motion then the flow is confined to a surface on which the integral is constant.
This can be seen most easily by transforming to angle-action coordinates, where the
surface is an invariant (multidimensional) torus.

A system with $n$~coordinates and
$n$~constants of the motion is \emph{integrable} and cannot have chaos
(though the motion can still be quasiperiodic or exhibit other complicated behavior).
For example, we can consider 
geodesic orbits around a Kerr black hole to have eight degrees of freedom
($n=4$) and four constants of the motion---particle rest mass~$\mu$, energy~$E$, axial or
$z$~angular
momentum~$L_z$, and Carter constant~$Q$---which 
are enough to integrate the equations of motion
explicitly. Alternatively, we may look at Kerr spacetime as having a 6-dimensional phase
space by eliminating time (which is simply a re-parameterization of the proper time)
and using rest mass conservation to eliminate one momentum coordinate.
Then the three integrals $E$,~$L_z$, and~$Q$ are sufficient to integrate the motion.
(In practice, we allow all four momenta to evolve freely; the normalization is then
a constraint which can be checked for consistency at the end of the integration.)

In the case of a spinning test particle, the extra spin degrees of freedom create
the possibility for chaotic behavior.  Moreover, since~$Q$ is not conserved in the
case of nonzero spin,
even without the extra spin degrees of freedom
the potential for chaos would exist.  Kerr spacetime 
has just enough constants to make the system integrable; 
losing~$Q$ reduces the number of analytic integrals below the critical level required
to guarantee integrability.\footnote{It is possible that deformations of Kerr geometry
that destroy~$Q$ nevertheless possess a numerical integral that preserves integrability, 
in analogy with some
galactic potentials~\citep{BinneyTremaine1987}, but 
the loss of~$Q$ certainly ends the \emph{guarantee} of integrability.}

\subsection{Hamiltonian systems}
\label{sec:Hamiltonian_systems}

\subsubsection{Lyapunov exponents for Hamiltonian flows}

The phase space flow of Hamiltonian systems is constrained
by more than the integrals of the motion.  In particular,
the Lyapunov exponents of a Hamiltonian system come in pairs
$\pm\lambda$; i.e., if $\lambda$ is a Lyapunov exponent then so is
$-\lambda$~\citep{EckmannRuelle1985}. Geometrically, this means that
if one semimajor axis of the phase-space ellipsoid stretches an amount
$e^{\lambda \tau}=L$, another axis must shrink by an amount $e^{-\lambda
\tau}=1/L$.  One consequence of this property is that the product of the
lengths of the axes is~1.  Since the ellipsoid volume is proportional
to this invariant product, Liouville's theorem on the conservation of
phase space volume follows as a corollary.

The $\pm\lambda$ property of Hamiltonian flows results from the symplectic nature of the
Jacobian matrix for Hamiltonian dynamical systems.\footnote{A matrix~$S$ is symplectic
with respect to the canonical symplectic matrix~$J$ if $S^T J S=J$, where 
$ J=\left(
\begin{array}{cc}
0 & -I\\
I & 0
\end{array}\right)$
and $I$ is the $n\times n$ identity matrix.}
But a naive analysis of the Jacobian matrix of the Papapetrou equations
shows that it is not symplectic
in the canonical sense.
Nevertheless, the Papapetrou equations
can be derived from a Lagrangian~\citep{Hojman1975}, and can be
cast in Hamiltonian form by use of a free Hamiltonian with added
constraints (following the method of Dirac~\citep{Dirac} as discussed
in~\citep{HansonRegge1974}). As a consequence, we could in principle find
coordinates in which the Jacobian matrix is symplectic with respect to the
canonical symplectic matrix. Fortunately, this is an unnecessary complication, since
the underlying dynamics are independent of the coordinates.

\subsubsection{Exponents for spinning test particles}

As discussed in Sec.~\ref{sec:modified_GS}, the lack of explicit time dependence
independence and the three constraints
reduce the degrees of freedom from twelve to eight, which leaves the possibility
of eight nonzero Lyapunov exponents.
The phase space flow is further constrained by the
constants of the motion,
energy and $z$~angular momentum;
corresponding to each
constant should be a zero Lyapunov exponent, since trajectories that
start on an invariant torus must remain there. 
This leaves six exponents potentially nonzero.
Since the exponents must come in pairs~$\pm\lambda$, there should be at
most three independent nonzero exponents.

\section{Results}
\label{sec:results}

First we give results for the dynamics of 
the Papapetrou equations in the extreme (and unphysical) limit~$S=1$,
which represents a violation of the test-particle approximation but is still
mathematically well-defined.  We find the presence of chaotic orbits (in agreement
with~\citep{SuzukiMaeda1997}).  We next examine the effects of varying~$S$, including the
limit~$S\ll1$. Finally,
we investigate more thoroughly the dynamics for physically realistic spins.

\subsection{Chaos for $S=1$}

\subsubsection{Maximally spinning Kerr spacetime}

In a background spacetime of a maximally spinning Kerr black hole ($a=1$)
there are
unambiguous positive Lyapunov exponents for a range of physical
parameters when~$S=1$.
We show a typical orbit that produces nonzero Lyapunov
exponents in Fig.~\ref{fig:kerr_spin_orbit}. The orbit has energy $E=0.8837\,\mu$,
$z$~angular momentum $J_z=2.0667\,\mu M$, and the radius ranges from pericenter
$r_p = 1.7\,M$
to apocenter $r_a = 6.7\,M$.  
The Lyapunov integrations typically run for $10^4\,M$,
which corresponds approximately to $400$~$\phi$-orbital periods.

\begin{figure*}
\begin{tabular}{ccc}
\includegraphics[width=2.in]{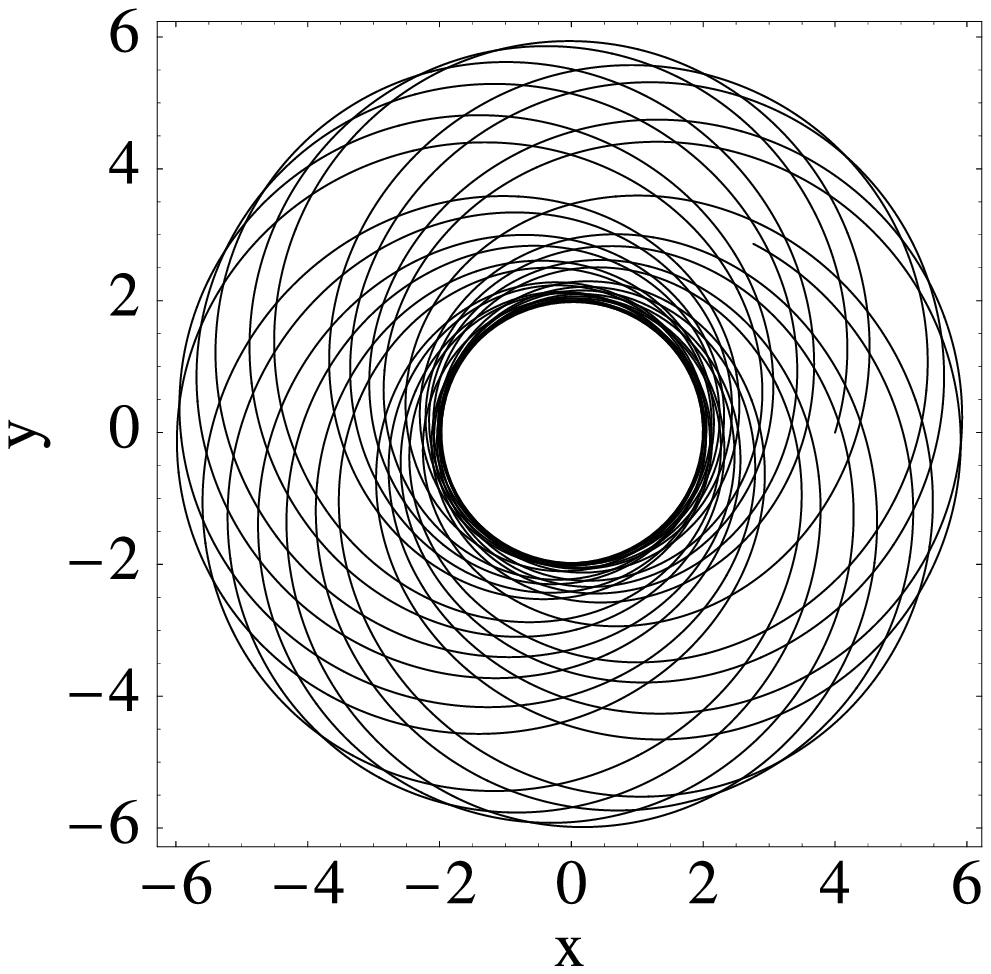} & \hspace{0.5in}
	& \includegraphics[width=3.in]{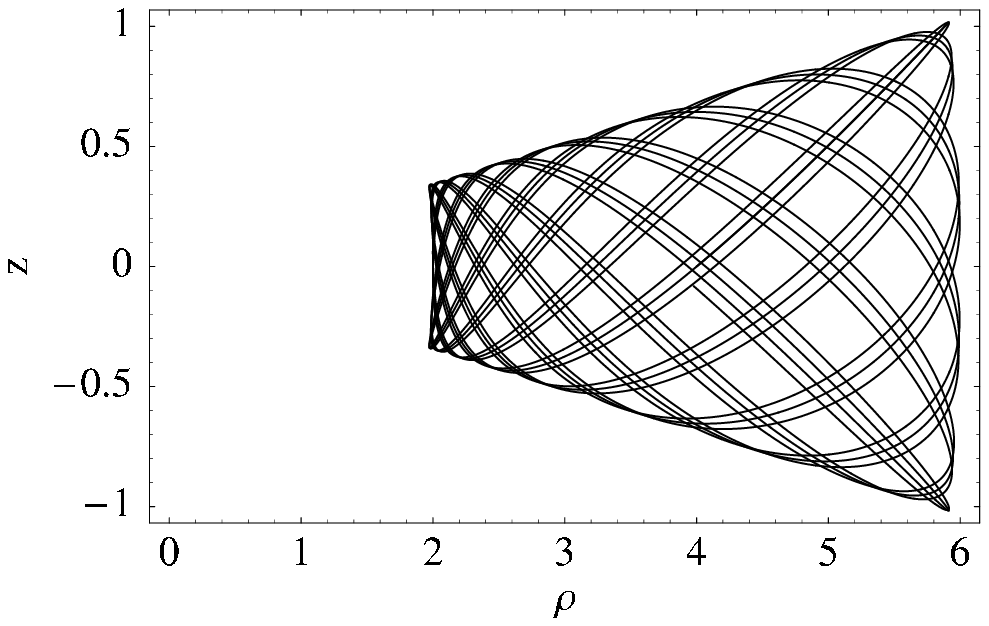}\\
(a) & & (b)\medskip\\
\end{tabular}
\caption{\label{fig:kerr_nonspin_orbit}
The orbit of a non-spinning~($S=0$)
test particle in maximal ($a=1$) Kerr spacetime, plotted in Boyer-Lindquist coordinates.
(a) $y=r\sin\theta\sin\phi$ vs.\ $x=r\sin\theta\cos\phi$; 
(b) $z$ vs.\ $\rho=\sqrt{x^2+y^2}$. 
The orbital parameters are $E=0.8837\,\mu$ and $J_z=2.0667\,\mu M$,
with pericenter $2.0\,M$ and apocenter $6.0\,M$. 
}
\end{figure*}

\begin{figure*}
\begin{tabular}{ccc}
\includegraphics[width=2.in]{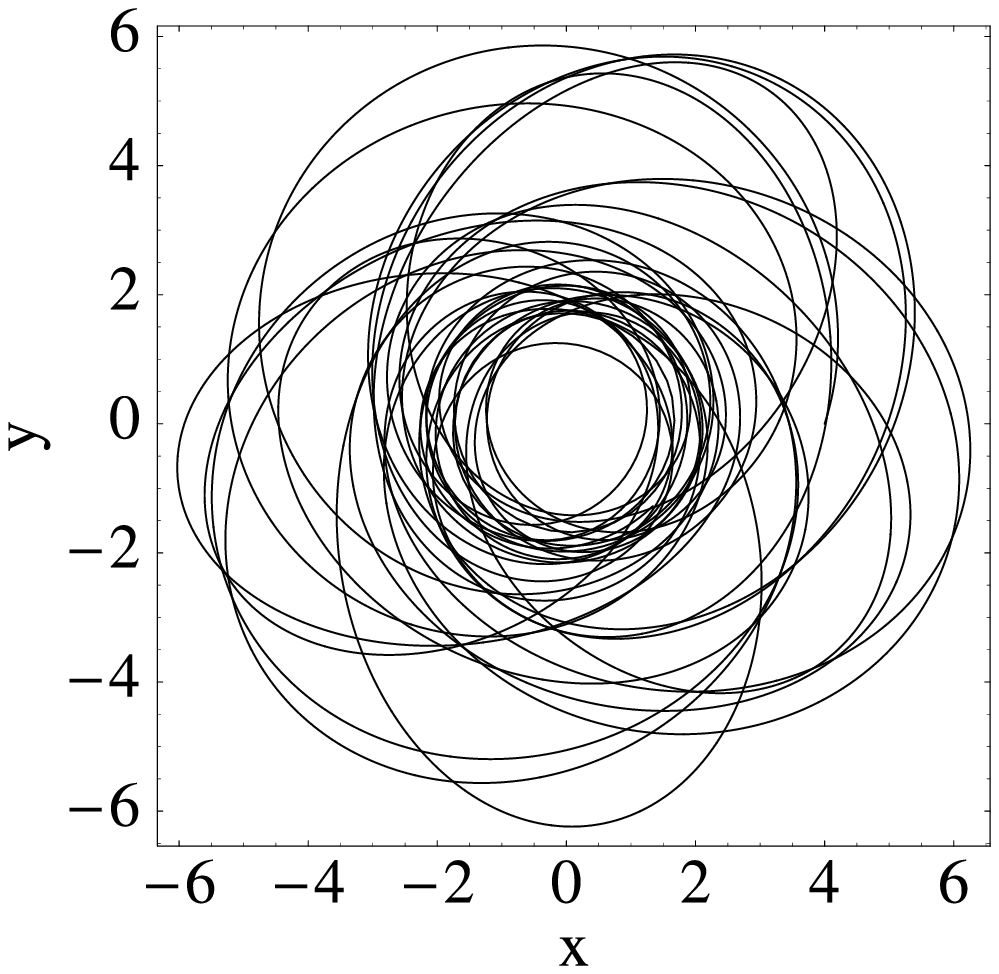} & \hspace{0.5in}
	& \includegraphics[width=3in]{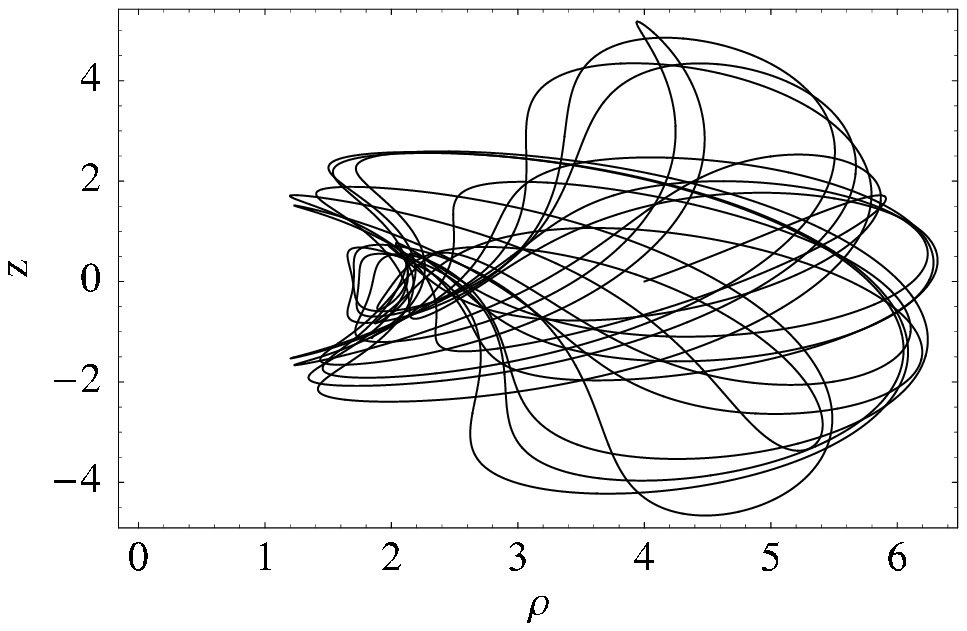}\\
(a) & & (b)\medskip\\
\end{tabular}
\caption{\label{fig:kerr_spin_orbit}
The orbit of a maximally spinning~($S=1$)
particle in maximal Kerr spacetime, for $E=0.8837\,\mu$ and $J_z=2.0667\,\mu M$ (the same
values as in Fig.~\ref{fig:kerr_nonspin_orbit} above).
The spin has initial
values of~$S^{\hat r} = S^{\hat\mu} = 0.1$, corresponding to an
initial angle of~$54^\circ$ with respect to the vertical
in the particle's rest frame.
As in Fig.~\ref{fig:kerr_nonspin_orbit}, 
we plot $y$ vs.\ $x$ in~(a) and $z$ vs.\ $\rho$ in~(b).
The spin causes significant deviations from geodesic orbits.}
\end{figure*}

\begin{figure*}
\begin{tabular}{ccc}
\includegraphics[width=3.in]{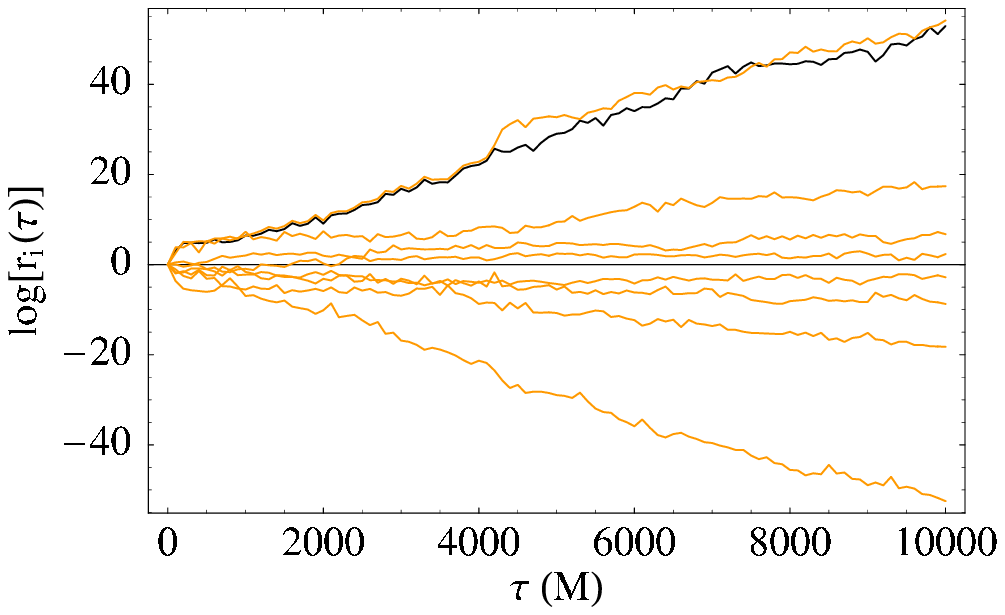} & \hspace{0.5in}
	& \includegraphics[width=3in]{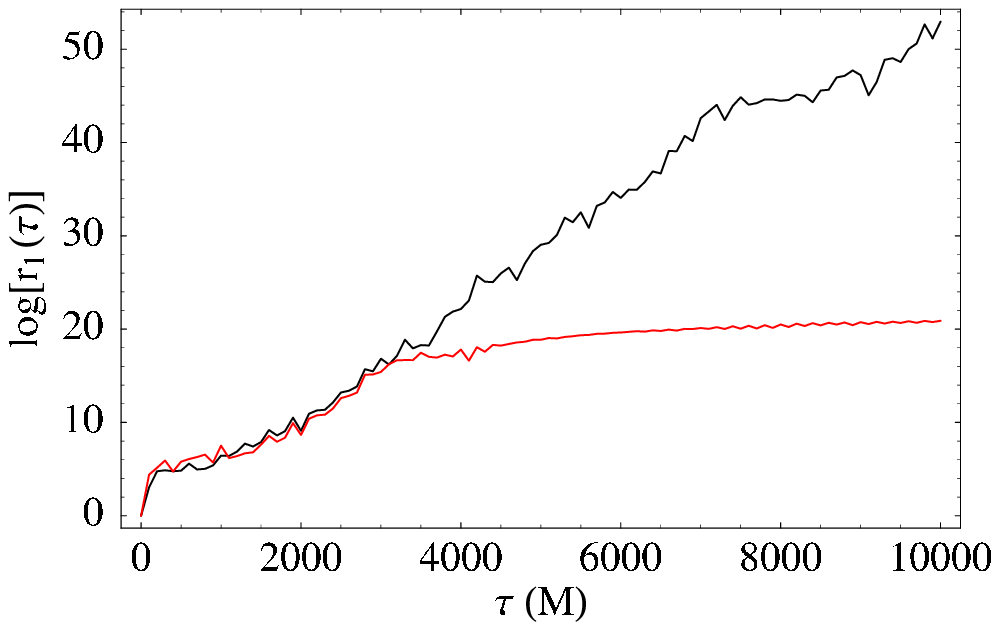}\\
(a) & & (b)\medskip\\
\end{tabular}
\caption{
\label{fig:kerr_lyap_evol}
Natural logarithms of the phase space ellipsoid axes vs.\ proper time in Kerr spacetime
with~$S=1$.
The slopes of the lines are the Lyapunov exponents; the largest exponent is approximately
$\lambda_\mathrm{max} = 5\times10^{-3}\,M^{-1}$.
The initial conditions
are the same as in Fig.~\ref{fig:kerr_spin_orbit}, and one point is recorded at
each time~$T = 100\,M$ (Sec.~\ref{sec:algorithm_detail}).
(a)~Full Gram-Schmidt Jacobian method (light) with rigorous Jacobian method (dark).
The full~GS method is rescaled at each time~$T$ according the algorithm in 
Sec.~\ref{sec:num_lyap}, while the rigorous Jacobian method is unrescaled.
The two methods agree closely on the value of the largest Lyapunov exponent.
(b)~Rigorous Jacobian method compared to
unrescaled deviation vector method. Note that the latter method, which started 
with a deviation of size~$10^{-7}$, saturates at~$\sim\!\!16$. This corresponds to a
growth of $e^{16}\approx9\times10^{6}$, which means that the separation has grown to 
a size of order unity.}
\end{figure*}

\begin{figure}
\includegraphics[width=3.in]{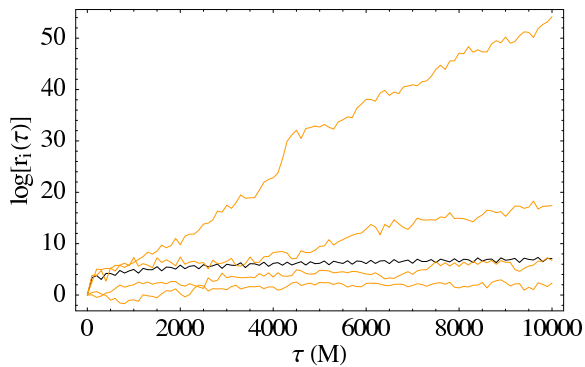}

\caption{Ellipsoid axis lengths from the upper half of
Fig.~\ref{fig:kerr_lyap_evol}(a)  (light), compared to an integration with zero
spin and hence zero Lyapunov exponent (dark).  Only two of the four lines
represent exponents distinguishable from zero.}

\label{fig:kerr_lyap_compare_zero}
\end{figure}

We can illustrate the presence of a chaotic orbit by plotting the natural logarithm
of the $i$th~ellipsoid axis $\log\,[r_i(\tau)]$
vs.\ $\tau$ [Eq.~(\ref{eq:log_r})], 
so that the slope is the Lyapunov exponent, as shown in
Fig.~\ref{fig:kerr_lyap_evol}.\footnote{It is traditional to
plot~$\log\,[r_i(\tau)]/\tau$, which converges to the Lyapunov exponent
as~$\tau\rightarrow\infty$, but it is much easier to identify the linear growth 
of~$\log\,[r_i(\tau)]$ than to identify the convergence 
of~$\log\,[r_i(\tau)]/\tau$.  The $\pm\lambda$ property is also clearer on such plots.}
There appear to be two nonzero Lyapunov exponents; the third
largest exponent is consistent with zero, as shown in Fig.~\ref{fig:kerr_lyap_compare_zero}.
The reflection symmetry of the figure is a consequence
of the exponent pairing: for each line with slope $\lambda$, there is a second line
with slope $-\lambda$.

The main plot in Fig.~\ref{fig:kerr_lyap_evol}(a) is generated by the modified Gram-Schmidt
algorithm (Sec.~\ref{sec:modified_GS}).  Recall that this method depends on the value
of~$\epsilon$ used to infer the tangent vector; we find a valid~$\epsilon$ by calibrating
it using the rigorous Jacobian method, which must yield an exponent that matches the
largest exponent from the modified Gram-Schmidt method. 
The plot in Fig.~\ref{fig:kerr_lyap_evol}(a) represents the case~$\epsilon=10^{-6}$;
it is apparent that the two methods agree closely.
The unrescaled deviation vector method provides
an additional check on the validity of the largest
exponent, as shown in Fig.~\ref{fig:kerr_lyap_evol}(b).  As expected, the unrescaled approach
closely tracks the full Jacobian approach until it saturates.

The numerical values of the exponents are shown in
Table~\ref{table:kerr_exponents}. The $\pm\lambda$ property is best satisfied
by $\pm\lambda_{\rm max}$, the exponents with the largest absolute value. The
exponents are least-squares fits to the data, with approximate standard errors
of~$1\%$.   These errors are not particularly meaningful since the exponents
themselves can vary by~$\sim\!\!10\%$ depending on the initial direction of the
deviation vector.  Moreover, even exponents that appear nonzero may be
indistinguishable from zero in the sense of
Fig.~\ref{fig:kerr_lyap_compare_zero}; for such exponents a ``$1\%$'' error on
the fit is meaningless.

For initial conditions considered in Fig.~\ref{fig:kerr_spin_orbit},
and other orbits in the strongly relativistic region near
the horizon, typical largest Lyapunov exponents are on the
order of a ${\rm few}\times10^{-3}/M$. For the particular case
illustrated in Fig.~\ref{fig:kerr_spin_orbit}, we have $\lambda_{\rm
max}\approx5\times10^{-3}\,M^{-1}$, which implies an $e$-folding
timescale of $\tau_\lambda\equiv1/\lambda\approx2\times10^2\,M$.
This is strongly chaotic, with a significant divergence in approximately
eight~$\phi$-orbital periods.

Based on integrations in the case of zero spin, which corresponds
to no chaos (Lyapunov exponents all zero), we can determine how
quickly the exponents approach zero numerically.\footnote{As
noted in the introduction, it is possible for integrable but
unstable orbits to have positive Lyapunov exponents.  We avoid
this issue by choosing a baseline orbit that is not unstable.}
Fig.~\ref{fig:kerr_lyap_compare_zero} compares the four apparently
positive exponents with a known zero exponent.  Only two of the four
exponents are unambiguously distinguishable from zero, consistent with
the argument in Sec.~\ref{sec:Hamiltonian_systems} that there should be
at most three independent nonzero exponents.

Finally, we note that the components of the direction of largest stretching are all
nonzero in general.  The chaos is not confined to the spin variables alone, but rather
mixes all directions.  This indicates that chaos could in principle manifest itself in
the gravitational waves from extreme mass-ratio binaries---but see
Sec.~\ref{sec:physically_realistic} below.

\begin{table}
\caption{\label{table:kerr_exponents}
Lyapunov exponents in Kerr spacetime
in units of $10^{-3}\,M^{-1}$, using a least squares fit.  The
exponents correspond to the semimajor axis evolution shown in Fig.~\ref{fig:kerr_lyap_evol}(a).
As is typical with the Gram-Schmidt
Jacobian method, the exponents with the largest magnitudes are determined most
accurately, and thus show the $\pm\lambda$ property most clearly.
The standard errors on the fit are $\sim\!\!1\%$ for each exponent, but these errors
are dominated by two systematic errors:
(1)~the variation due to different choices
of initial (random) tangent vectors; (2)~nonzero numerical values even for
exponents that converge to zero eventually.  In particular, the four
smallest exponents (in absolute value) are indistinguishable from zero
(see Fig.~\ref{fig:kerr_lyap_compare_zero}).}
\begin{ruledtabular}
\begin{tabular}{ccccc}
$+\lambda$ & $5.5$ & $1.5$  & $0.56$  & $0.25$ \\
$-\lambda$ & $5.3$ & $1.6$ & $0.76$ & $0.072$
\end{tabular}
\end{ruledtabular}
\end{table}

\subsubsection{Schwarzschild spacetime revisited}

\begin{figure*}
\begin{tabular}{ccc}
\includegraphics[width=2.in]{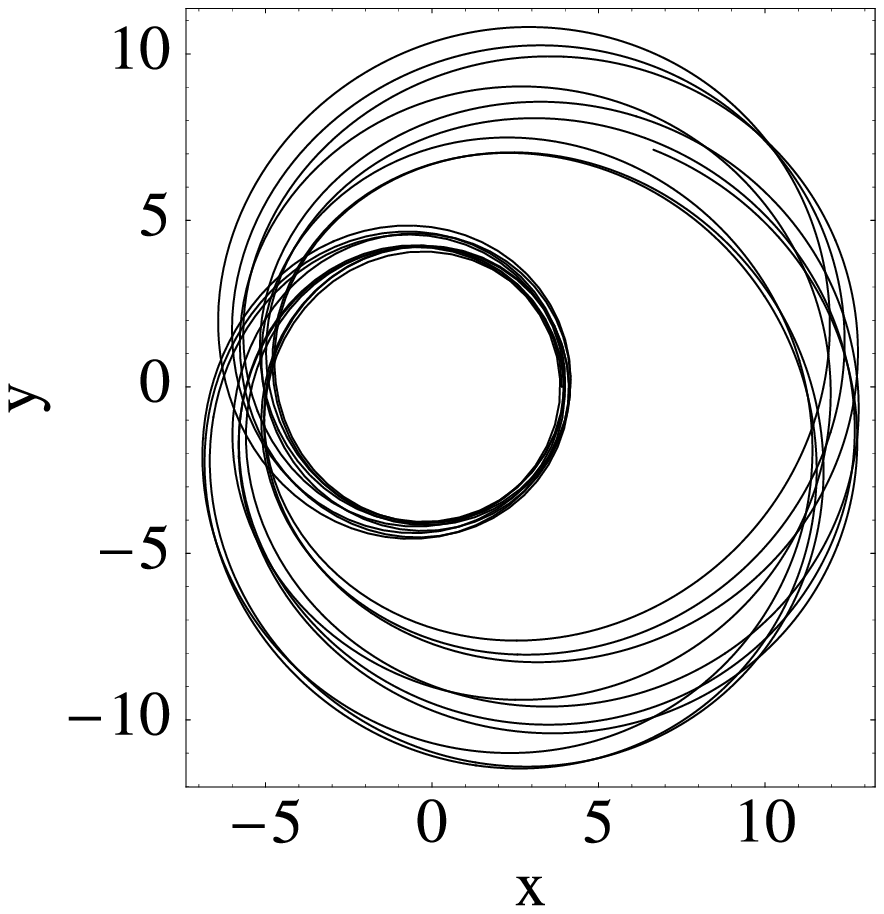} & \hspace{0.5in}
	& \includegraphics[width=3.in]{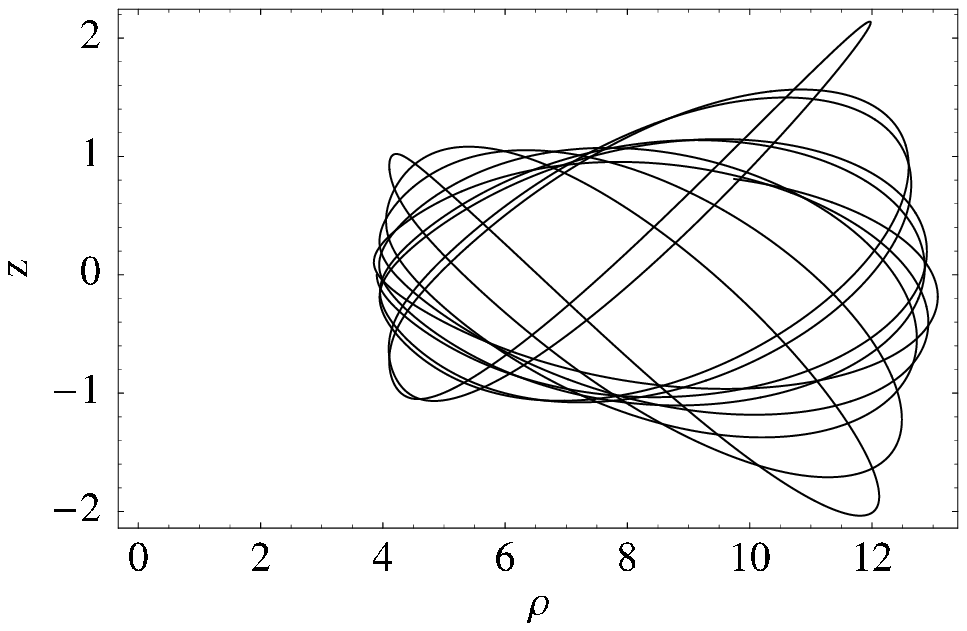}\\
(a) & & (b)\medskip\\
\end{tabular}
\caption{\label{fig:schw_spin_orbit}
The orbit of a maximally spinning~($S=1$)
test particle in Schwarzschild spacetime for $E=0.94738162\,\mu$ and $J_z=4.0\,\mu M$
As before, we plot (a)~
$y$ vs.\ $x$ and~(b) $z$ vs.\ $\rho=\sqrt{x^2+y^2}$.}
\end{figure*}

\begin{figure*} 
\begin{tabular}{ccc}
\includegraphics[width=3.in]{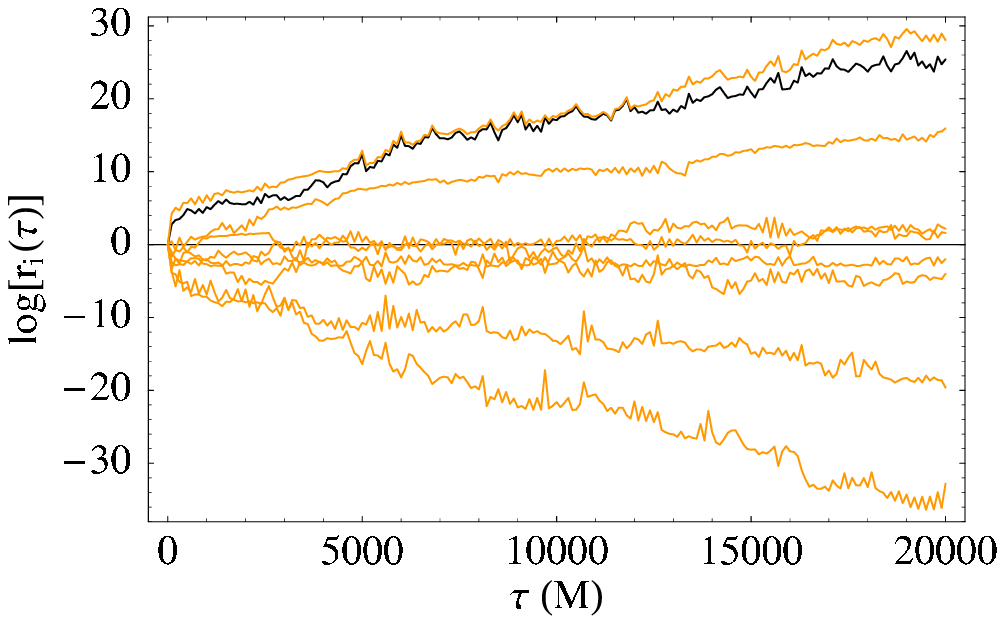} & \hspace{0.5in}
	& \includegraphics[width=3in]{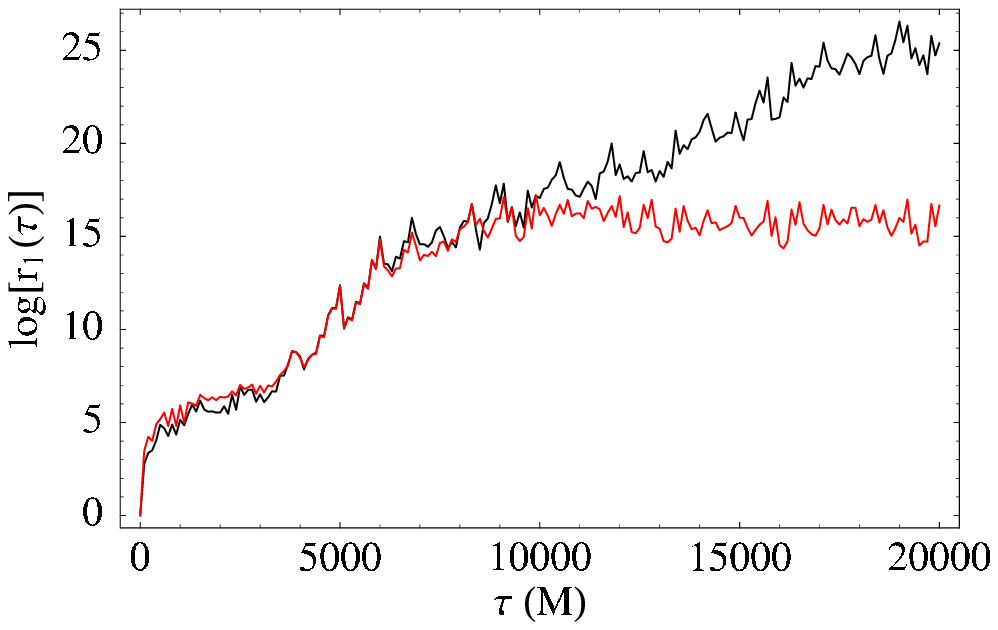}\\
(a) & & (b)\medskip\\
\end{tabular}
\caption{\label{fig:schw_lyap_evol}
Natural logarithms of the phase space ellipsoid axes vs.\ proper time in
Schwarzschild spacetime with~$S=1$. The largest exponent is
$\lambda_\mathrm{max} \approx 1.2\times10^{-3}\,M^{-1}$. The initial conditions
are the same as in Fig.~\ref{fig:schw_spin_orbit}. (a)~Full Gram-Schmidt
Jacobian method (light) with rigorous Jacobian method (dark). (b)~Rigorous
Jacobian method compared to unrescaled deviation vector method.   As
in~Fig.~\ref{fig:kerr_lyap_evol}(b), the unrescaled method eventually
saturates.}
\end{figure*}

We now reconsider the case of a spin~$S=1$ particle in Schwarzschild
spacetime, as investigated in Ref.~\citep{SuzukiMaeda1997}.
Fig.~\ref{fig:schw_spin_orbit} shows an orbit similar to a chaotic
orbit considered there (Fig.~4(d) in~\citep{SuzukiMaeda1997}).  A plot
of $\log[r_i(\tau)]$ vs.\ $\tau$ (Fig.~\ref{fig:schw_lyap_evol})
shows behavior similar to that in Fig.~\ref{fig:kerr_lyap_evol}.
In particular, the $\pm\lambda$ symmetry is present, apparently with two
positive exponents.  (The other lines are indistinguishable
from zero, again using $S=0$ orbits as a baseline.)  The largest
exponent of~$1.5\times10^{-3}\,M^{-1}$ agrees closely with the
value from Ref.~\citep{SuzukiMaeda1997}, which reported an exponent
of~$\sim\!\!2\times10^{-3}\,M^{-1}$ for a similar orbit. (This agreement
is somewhat surprising, since~\cite{SuzukiMaeda1997}
appears not to have taken the constrained nature
of the deviation vectors into account.
Luckily, the exponents are robust, and even unconstrained deviation vectors give nearly
correct results.)

\begin{table}
\caption{\label{table:schw_exponents}
Lyapunov exponents in Schwarzschild spacetime in units of $10^{-3}\,M^{-1}$,
using a least squares fit.  The exponents correspond to the semimajor axis
evolution shown in Fig.~\ref{fig:schw_lyap_evol}(a), which is similar to the
orbit in Fig.~4(d) of Ref.~\citep{SuzukiMaeda1997}. As with the Kerr case
(Table~\ref{table:kerr_exponents}),  the standard errors on the fit are
$\sim\!\!1\%$ for each exponent, and the same caveats apply.
The four smallest exponents (in absolute vale) 
are indistinguishable from zero in the sense of
Fig.~\ref{fig:kerr_lyap_compare_zero}.}
\begin{ruledtabular}
\begin{tabular}{ccccc}
$+\lambda$ & $1.2$ & $0.67$  & $0.21$  & $0.0063$ \\
$-\lambda$ & $1.5$ & $0.57$ & $0.10$ & $0.00023$
\end{tabular}
\end{ruledtabular}
\end{table}

\subsubsection{Kerr and Schwarzschild compared}
\label{sec:kerr_schw}

The Kerr and Schwarzschild Lyapunov exponents of the previous two sections are not all
that different; both are $10^{-2}$--$10^{-3}\,M^{-1}$ in order of magnitude.  Nevertheless, the
two systems prove to be quite different: chaotic orbits are easy to find in Kerr
spacetime for nearly any initial condition that explores the strongly relativistic region
near the horizon, whereas nearly all
analogous orbits in Schwarzschild spacetime are not chaotic.

\begin{figure}
\includegraphics[width=3in]{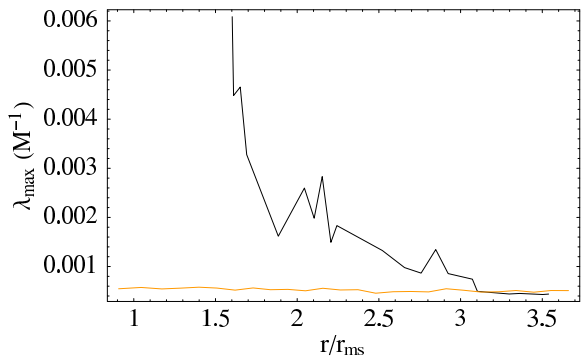}
\caption{\label{fig:kerr_vs_schw}
Comparison of maximally spinning ($S=1$) particle orbits in
Kerr (dark) and Schwarzschild (light). We plot the largest Lyapunov exponent
versus pericenter (normalized by the marginally stable radius).  The Kerr initial
conditions for the innermost orbits are essentially as 
in~Fig.~\ref{fig:kerr_spin_orbit}. The Schwarzschild orbits are identical to their Kerr
counterparts in
inclination~($10^\circ$) and eccentricity~($e=0.5$) but have the Kerr parameter~$a$ set
to zero.  The Schwarzschild orbits have exponents indistinguishable from zero over the
entire range of parameters.}
\end{figure}

Fig.~\ref{fig:kerr_vs_schw} compares Kerr and Schwarzschild orbits with the same
inclination angle~$\iota=10^{\circ}$ and eccentricity~$e=0.5$ 
but varying pericenters~$r_p$.
(Details of this parameterization method, mentioned above
in Sec.~\ref{sec:parameterization}, appear in~\citep{Hartl_2_2002}.)
We insure that the systems are analogous by using orbits
of~$S=1$ particles with the same values of $r_p/r_{\mathrm{ms}}$, where $r_{\mathrm{ms}}$ 
is the radius of the marginally stable orbit 
in the corresponding~$S=0$ (geodesic) case. 
We use a Kerr geodesic integrator 
developed by Scott Hughes~\citep{HughesIntegrator}
to find $r_{\mathrm{ms}}$, which is the smallest pericenter that still yields a stable orbit. 
For the values of $\iota$~and~$e$ considered, $r_{\mathrm{ms}}=1.0\,M$ for Kerr and
$r_{\mathrm{ms}}=4.67\,M$ for Schwarzschild.

It is evident from Fig.~\ref{fig:kerr_vs_schw} that the Kerr orbits are chaotic for a
broad range of pericenters, with the maximum Lyapunov $\lambda_{\mathrm{max}}$ generally
decreasing as the pericenter increases.  In contrast, the Schwarzschild orbits are not
chaotic anywhere over the entire range of valid initial conditions.  In fact, we are unable to
find any chaotic orbits in Schwarzschild spacetime
other than the types identified by Suzuki and Maeda~\cite{SuzukiMaeda1997},
which were exceptional cases of orbits on the edge of a generalized effective potential.
In Kerr, on the other hand, chaotic orbits appear to be the rule for pericenters 
near~$r_{\mathrm{ms}}$.

\subsection{Dependence on $S$}
\label{sec:spin_dependence}

\begin{figure*}
\begin{tabular}{ccc}
\includegraphics[width=3in]{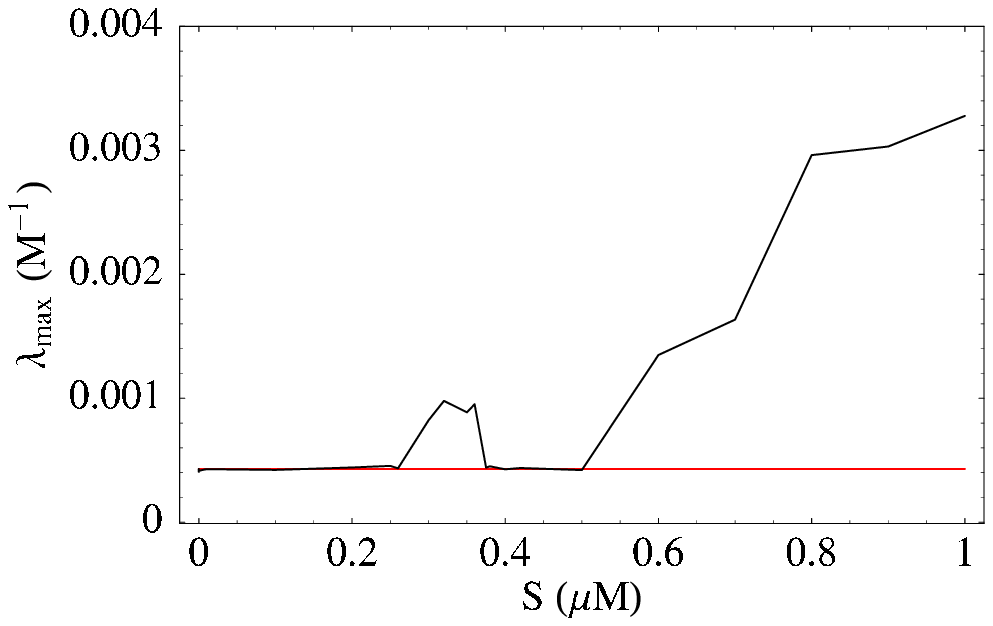} & \hspace{0.5in}
	& \includegraphics[width=3in]{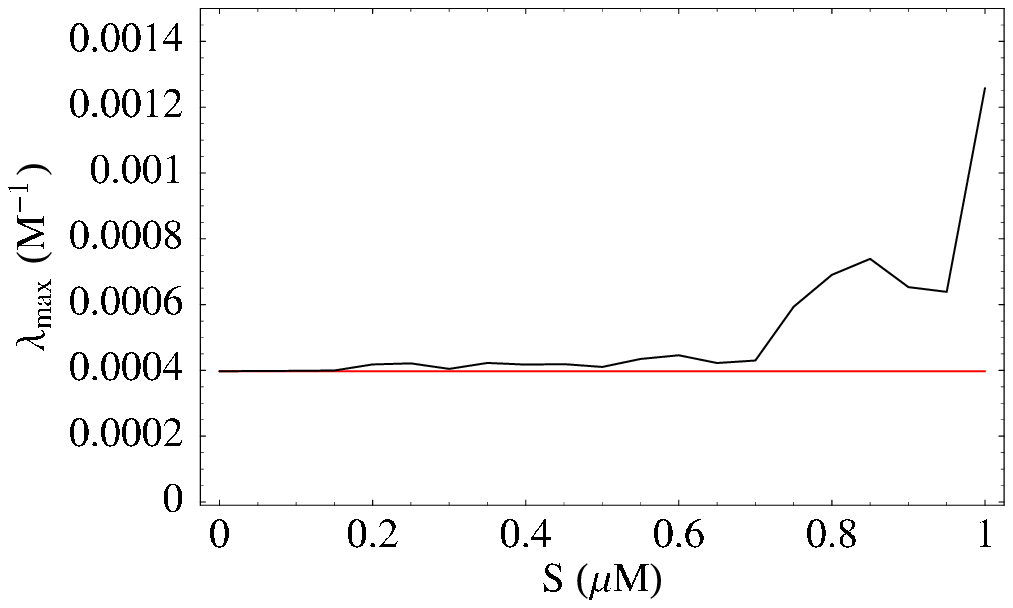}\\
(a) & & (b)\medskip\\
\end{tabular}

\caption{\label{fig:variation_S} Variation of largest Lyapunov exponent
vs.~$S$. (a)~The spin~$S=1$ initial conditions are the same as in
Fig.~\ref{fig:kerr_spin_orbit}. (b)~Another $S=1$ case with a different
inclination angle~($20^\circ$) and pericenter~($2.5\,M$). As the spin
decreases, we hold fixed the Kerr orbital parameters: inclination angle,
eccentricity, and pericenter. Note that in (a) the chaos disappears below
$S\sim0.5$, but returns in a region centered on $S\sim0.3$.  The horizontal line
in both plots is the value of~$\lambda_\mathrm{max}$ calculated for the
baseline~$S=0$ orbit.
In both (a)~and~(b)
the Lyapunov exponent is indistinguishable from zero for physically realistic
spins.} \end{figure*}

Since chaos must disappear as $S\rightarrow0$, we expect to see the largest Lyapunov
exponent approach zero in this limit.
This is indeed the case: in Fig.~\ref{fig:variation_S}, which shows the variation of
$\lambda_\mathrm{max}$ with $S$ for two different orbits, we see that the chaos 
unambiguously present when $S=1$ is not present for smaller values of~$S$.
In particular, the largest
Lyapunov exponent is indistinguishable from zero over the entire range $10^{-6}\leq
S\leq10^{-1}$. (The far left of the plots have data points for each decade
in this range.)

\begin{figure}
\includegraphics[width=3in]{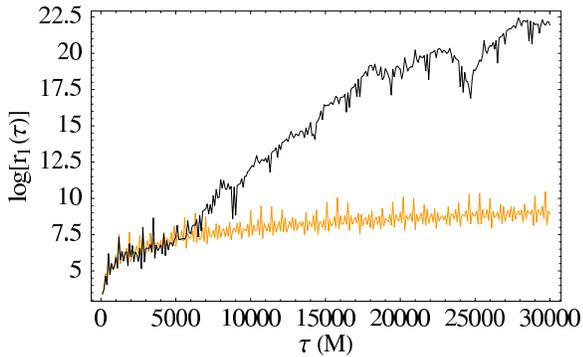}
\caption{\label{fig:bump_S}
Two orbits from the ``bump'' in Fig.~\ref{fig:variation_S}(a). The $S=0.4$ orbit (light) is
not chaotic, but the $S=0.3$ orbit (dark) is chaotic, despite having a smaller value of
the spin.}
\end{figure}

Although the strength of the chaos generally decreases with~$S$, one remarkable
feature of Fig.~\ref{fig:variation_S}(a) is the return of chaotic orbits
between $S\sim0.25$ and $0.4$ after their disappearance at $S\sim0.5$.  The
effect is qualitatively clear in Fig.~\ref{fig:bump_S}.  This chaotic ``bump''
in $\lambda_\mathrm{max}$~vs.~$S$ illustrates an important theme in nonlinear
dynamical systems: the \emph{only} way to determine whether an orbit is chaotic
is to do the calculation.  Though we certainly expect the strength of chaos to
be smaller for $S\ll 1$ than for $S\approx 1$,  it is impossible, in
general, to determine \emph{a priori} whether a particular set of parameters
will lead to chaotic behavior.

\subsection{Physically realistic spins}
\label{sec:physically_realistic}

The Papapetrou equations are only realistic in the test-particle limit, so
physically realistic spins must satisfy~$S\ll1$ (Sec.~\ref{sec:spin_param}).
This corresponds to likely sources of gravitational waves for 
LISA~\citep{FinnThorne2000,Hughes2000,Hughes2001_2}, e.g.,
maximally spinning $\mu=10\,M_\odot$ black holes spiraling into supermassive
$M=10^6\,M_\odot$ Kerr black holes, which have
spin parameters of $S=\mu/M=10^{-5}$.  Because of their likely importance as emitters of
gravitational waves, it is essential to understand the dynamics of
such systems.

\subsubsection{Vanishing Lyapunov exponents}

We would like to be able to make a definitive statement about the presence or
absence of chaos for ``small'' spins, e.g., values of~$S$ in the
range~$10^{-2}$--$10^{-6}$. Unfortunately, when determining Lyapunov exponents
numerically, it is impossible to conclude definitively that an orbit is or is
not chaotic,  since to do so would require an infinite-time integration. On the
other hand, for suspected non-chaotic orbits, we can provide an approximate
bound on the $e$-folding timescale. 

\begin{figure}
\includegraphics[width=3in]{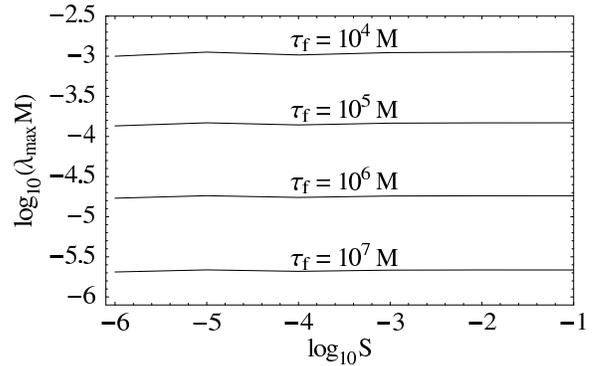}

\caption{\label{fig:S_zero} The variation of the dimensionless 
quantity~$\lambda_\mathrm{max}\,M$ with final integration time~$\tau_f$ for spin
parameter~$S$ in the range $10^{-2}\leq S\leq10^{-6}$.  From top to bottom, the
total integration time varies from $10^4\,M$ to $10^7\,M$.  It is likely that
the true Lyapunov exponent is zero. }

\end{figure}

The numerical values of exponents suspected to be zero
depend strongly on the time of the integration.
For example, for values of~$S$ in the range $10^{-2}\leq S\leq10^{-6}$,
the exponent in Fig.~\ref{fig:variation_S} 
appears to be~$\lambda_\mathrm{max}\approx5\times10^{-4}\,M^{-1}$, but this plot
represents an integration time of only~$10^4\,M$.  
Longer integration times give correspondingly smaller estimates for the
suspected zero exponents (Fig.~\ref{fig:S_zero}).
For the system shown in Fig.~\ref{fig:variation_S},
an integration of~$10^7\,M$ yields an estimate of 
$\lambda_\mathrm{max}\approx3.0\times10^{-7}\,M^{-1}$
for all spins in the range $10^{-2}\leq S\leq10^{-7}$.  In this case,
the relevant Lyapunov timescales are
at least $3\times10^{6}\,M$, and are probably much longer; the size of the bound is
limited only by our patience and computer budget.
It seems highly likely that such orbits are not
chaotic.

\subsubsection{Spin-induced phase differences}

\begin{table*}
\caption{\label{table:delta-phi_a=0.5}
Phase shifts~$\Delta\phi = \phi_\mathrm{geodesic} - \phi_\mathrm{spin}$ in
radians as a function of orbital inclination angle~$\iota$ and pericenter~$r_p$
for $a=0.5$ and $S=10^{-5}$. Inclination angle~$\iota=0^\circ$ is prograde
equatorial and $\iota=180^\circ$~is retrograde equatorial. The geodesic orbits
and their corresponding spin orbits start with the same initial
4-velocity~$v^{\mu}$, and the integrations are performed using Boyer-Lindquist
coordinate time~$t$, with $t_\mathrm{max} \approx$ (2000 times the average 
radial orbital period). The pericenters are scaled by the marginally stable
radius~$r_\mathrm{ms}$, and we start at $r_p/r_\mathrm{ms}=1.5$~to guarantee
the existence of valid initial conditions for the non-geodesic orbit. The spin
has fixed initial values of~$S^{\hat r} = S^{\hat\mu} = 0.1\,S$ (with hats
indicating an orthonormal basis), corresponding to initial angles of~$9^\circ$
to~$30^\circ$ with respect to the vertical in the particle's rest frame,
increasing with decreasing pericenter.}
\begin{ruledtabular}
\begin{tabular}{c|ccccccccc}
\multicolumn{9}{c}{$r_p/r_\mathrm{ms}$} \\
\hline
$\iota$ & 1.5 & 2.0 & 2.5 & 3.0 & 3.5 & 4.0 & 4.5 & 5\\
\hline
$10^\circ$ & 1.50e-02 & 5.69e-03 & 4.32e-03 & 2.13e-03 & 2.02e-03 & 1.27e-03 & 1.14e-03 & 7.77e-04 \\
$45^\circ$ & 2.79e-02 & 1.23e-02 & 1.01e-02 & 4.34e-03 & 4.60e-03 & 2.24e-03 & 1.66e-03 & 1.83e-03 \\
$85^\circ$ & 4.36e-02 & 2.92e-03 & 1.48e-03 & 8.24e-04 & 1.00e-03 & 2.20e-03 & 1.86e-03 & 1.26e-03 \\
$135^\circ$ & -9.02e-03 & -6.25e-03 & -2.34e-03 & -1.30e-03 & -1.73e-03 & -8.17e-04 & -6.76e-04 & -7.72e-04 \\
$170^\circ$ & 8.40e-04 & 2.85e-04 & 1.84e-04 & 7.31e-05 & 1.25e-04 & 1.12e-04 & 3.35e-05 & 3.07e-05 \\
\end{tabular}
\end{ruledtabular}
\end{table*}

\begin{table*}
\caption{\label{table:delta-phi_a=1}
Phase shifts~$\Delta\phi = \phi_\mathrm{geodesic} - \phi_\mathrm{spin}$ in radians as a function
of orbital inclination angle~$\iota$ and pericenter~$r_p$ for $a=1$ and $S=10^{-5}$.  
As in Table~\ref{table:delta-phi_a=0.5}, 
the pericenters are scaled by~$r_\mathrm{ms}$, and
the spin has fixed initial
values of~$S^{\hat r} = S^{\hat\mu} = 0.1\,S$ (corresponding in this case
to initial angles of~$28^\circ$ to~$61^\circ$, again decreasing
with increasing pericenter).}
\begin{ruledtabular}
\begin{tabular}{c|ccccccccc}
\multicolumn{9}{c}{$r_p/r_\mathrm{ms}$} \\
\hline
$\iota$ & 1.5 & 2.0 & 2.5 & 3.0 & 3.5 & 4.0 & 4.5 & 5 \\
\hline
$10^\circ$ & 7.21e-02 & 4.58e-02 & 2.41e-02 & 1.83e-02 & 1.10e-02 & 9.46e-03 & 6.56e-03 & 7.43e-03 \\
$45^\circ$ & 2.37e-01 & 5.56e-02 & 2.59e-02 & 1.83e-02 & 1.73e-02 & 1.52e-02 & 1.08e-02 & 7.83e-03 \\
$85^\circ$ & 1.96e-02 & 6.21e-03 & 2.82e-03 & 2.13e-03 & 2.66e-02 & 3.64e-03 & 6.47e-04 & 3.48e-03 \\
$135^\circ$ & -1.04e-02 & -3.17e-03 & -3.21e-03 & -1.41e-03 & -1.12e-03 & -8.46e-04 & -8.82e-04 & -5.59e-04 \\
$170^\circ$ & 3.89e-04 & 1.48e-04 & 6.68e-05 & 5.97e-05 & 8.09e-05 & 9.55e-05 & 3.06e-05 & 1.66e-05 \\
\end{tabular}
\end{ruledtabular}
\end{table*}

Even if their Lyapunov exponents are zero,
small spins affect the relative phase of the orbits, 
and since phase differences accumulate secularly~\citep{ACST1994}, 
the spin can still affect the gravitational wave signal.
It is therefore 
useful to have a sense of the orders of magnitude of such spin-induced phase-shifts.
Tables~\ref{table:delta-phi_a=0.5} and~\ref{table:delta-phi_a=1} show typical values for
the phase difference~$\Delta\phi = \phi_\mathrm{geodesic} - \phi_\mathrm{spin}$ for
$S=10^{-5}$, where the geodesic and spin systems start with the same initial
4-velocity~$v^\mu$.  The most useful quantity in practice is the phase shift as
measured by observers at infinity, so
we integrate in terms of the Boyer-Lindquist 
coordinate time~$t$ in place of~$\tau$. (This involves multiplying
the differential equations by $d\tau/dt$ at each time step.)
As is apparent from the tables, 
the phase shifts range broadly, from ~$10^{-1}$ to~$10^{-5}$ radians after
2000~radial orbital periods, but tend to decrease in magnitude with increasing
inclination angle or pericenter. 

Ref.~\cite{Hughes2001_2} shows that the number of orbital periods in a full 
inspiral from $r\approx4\,M$~to the final plunge is
$N\sim\frac{M}{\mu}$, which is $10^5$ for the systems in 
Tables~\ref{table:delta-phi_a=0.5} and~\ref{table:delta-phi_a=1}.
Since the table represents values of~$\Delta\phi$ 
for $2000$~times the average radial orbital period,
this means that the total phase shift during the inspiral is 
$50\,\Delta\phi_\mathrm{table}$.
For a $10^\circ$~inclination angle the total phase shift is on the order of a 
tenth of a radian to a radian.  Slightly more realistic values 
of the number of orbits can be obtained using
Fig.~2 in~\cite{Hughes2001_2}, which gives~$N\sim2\times10^4$ orbital periods
from $r=4\,M$ to the plunge at~$r\approx M$ for~$a=0.998$, 
$\iota=10^\circ$, and $M/\mu = 10^5$.  Since the orbit spends most of its
time between $4\,r_\mathrm{ms}$ and $2\,r_\mathrm{ms}$, interpolating in
Table~\ref{table:delta-phi_a=1} gives
$\Delta\phi_\mathrm{total}\approx10\times\Delta\phi_{r=3.0} = 2\times10^{-2}$.  
This is only a rough estimate, since the orbits in~\cite{Hughes2001_2} are circular,
while the orbits we consider are eccentric.

\subsection{Comments on time, rescaling, and norms}
\label{sec:comments}

In this paper, we have elected to use~$\tau$ as the time parameter,
a rescaling time~$T$ of $100\,M$, and a projected norm~(Sec.~\ref{sec:projnorm}).  
Here we discuss the effects of varying these choices.

\begin{figure}
\includegraphics[width=3in]{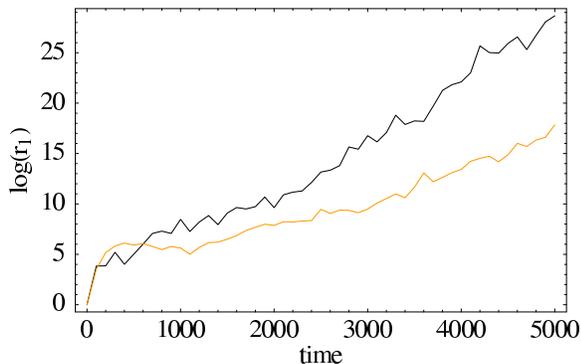}
\caption{\label{fig:taut}
The natural logarithm of the largest ellipsoid
axis vs.\ time for proper time~$\tau$ (dark) and coordinate time~$t$ (light).
The exponents are clearly different,
but the Lyapunov timescales~$\tau_\lambda = 1/\lambda_\tau$ 
and~$t_\lambda=1/\lambda_t$
are related by Eq.~(\ref{eq:taut}).}
\end{figure}

First, we consider the effects of using coordinate time~$t$ in place of~$\tau$.  In
Fig.~\ref{fig:taut}, we plot the natural logarithm of the largest ellipsoid
axis $\log[r_1(\tau)]$ vs.~$\tau$ together with $\log[r_1(t)]$ vs.~$t$.
(We use the unrescaled deviation vector approach for simplicity, since the Jacobian
approach requires a new Jacobian matrix for each coordinate
change.) The exponents are $\lambda_\tau = 5.05\times10^{-3}\,M^{-1}$ and 
$\lambda_t = 2.51\times10^{-3}\,M^{-1}$, implying Lyapunov timescales of 
$\tau_\lambda = 1.98\times10^2\,M$ and $t_\lambda = 3.98\times10^2\,M$.  The average
value of $dt/d\tau$ over the orbit is 2.06, whereas $t_\lambda/\tau_\lambda = 2.01$,
so the relationship
\begin{equation}
\label{eq:taut}
\frac{t_\lambda}{\tau_\lambda} = \left\langle \frac{dt}{d\tau}\right\rangle
\end{equation}
discussed in the introduction is well satisfied.

\begin{figure}
\includegraphics[width=3in]{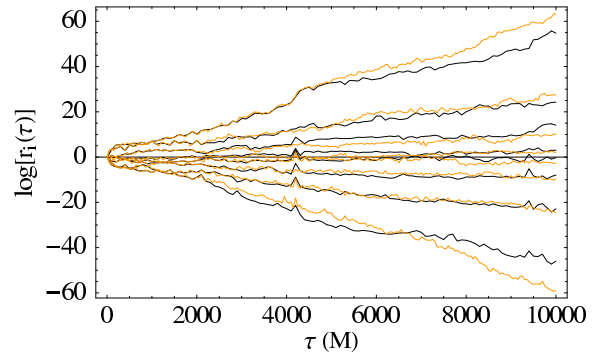}
\caption{\label{fig:timeT}
Natural logarithms of the ellipsoid
axes vs.~$\tau$ for rescaling time~$T=100\,M$ (dark) and time~$T=50\,M$ (light).
}
\end{figure}

Second, we discuss the effects of varying the rescaling time~$T$.  We find that
choosing~$T$ to be a moderate fraction of the shortest Lyapunov timescale
(corresponding to the largest Lyapunov exponent) works best, giving each axis
enough time to grow before rescaling while still keeping the negative exponents
from underflowing and preventing the largest axis from dominating. Rescaling
times between~$50\,M$ and~$100\,M$ work best for the systems we consider, which
have Lyapunov timescales ranging from $10^2\,M$ to~$10^3\,M$.   A comparison of
results for $T=50\,M$ and $T=100\,M$ appears in Fig.~\ref{fig:timeT}.

\begin{figure}
\includegraphics[width=3in]{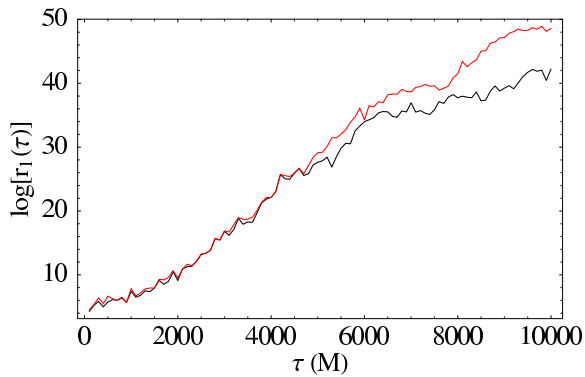}
\caption{\label{fig:euc_proj_compare}
The natural logarithm of the largest ellipsoid
axis vs.~$\tau$ for the Euclidean norm (top) and the projected norm
from Sec.~\ref{sec:projnorm} (bottom).
}
\end{figure}

Third, we compare the projected norm used here to a naive Euclidean norm for determining
the length of the phase-space tangent vectors~$\bm{\xi}_i$.  
As shown
in Fig.~\ref{fig:euc_proj_compare}, even using a 12-dimensional Euclidean norm changes
the resulting exponent very little (approximately 15\% in this example). 
Given its conceptual advantages, we choose to use the projected norm with the confidence
that the Lyapunov exponent order of magnitude is robust.

\section{Conclusions}

A spinning test particle, as described by the Papapetrou equations,
appears to be
chaotic in Kerr spacetime, 
with maximum $e$-folding timescales of a
$\textrm{few}\times10^2\,M$.
The applicability of this result is limited by three main 
factors: (1) chaos appears only for physically unrealistic values of the spin 
parameter; (2) other effects, such as tidal coupling, may be important for some
astrophysical systems, violating the pole-dipole approximation implicit in the
Papapetrou equations; and (3) we neglect gravitational radiation. The third limitation
is not fatal, since the radiation timescales can be long enough that chaos, if present
in the conservative limit,
would have time to manifest itself in the gravitational radiation of
extreme mass-ratio systems. 

In the unphysical~$S=1$ limit, the Lyapunov exponents exhibit characteristics
expected of a Hamiltonian system, appearing in pairs $\pm\lambda$
(Sec.~\ref{sec:Hamiltonian_systems}).  There are zero Lyapunov exponents which
correspond to the constants of the motion, but the other exponents are  in
general nonzero. (For the Kerr orbits considered in this paper,  we find that
two of the three independent exponents are nonzero, as illustrated in
Fig.~\ref{fig:kerr_lyap_compare_zero}.) Typical orders of magnitude for the
largest Lyapunov exponents are a~$\textrm{few}\times10^{-3}\,M^{-1}$ for
unphysical spins ($S=1$).   For physically realistic spin parameters
(Sec.~\ref{sec:physically_realistic}),  we find that
$\lambda_\mathrm{max}\alt\textrm{few}\times10^{-7}\,M^{-1}$, corresponding to
$e$-folding timescales of a~$\textrm{few}\times10^{6}\,M$. Even this bound 
appears to be limited only by the total integration time; in all physically
realistic cases considered, $\lambda_\mathrm{max}$ is indistinguishable from
zero (using $S=0$ integrations  as a baseline). 

From the perspective of gravitational radiation detection, our most important 
conclusion is that chaos seems to disappear
for physically realistic values of~$S$, i.e., values of~$S$ for which the test-particle
approximation and hence the Papapetrou equations are valid.
We are unable to comment on the dynamics of comparable mass-ratio binaries, since such
systems are not accurately modeled by the Papapetrou equations,
but for \emph{extreme} mass-ratio binaries
it appears unlikely that chaos will present a problem for the calculation of theoretical
templates for use in matched filters. 
A more thorough exploration of parameter space is
needed to reach a firmer conclusion~\citep{Hartl_2_2002}.

\begin{acknowledgments} I would like to thank Sterl Phinney for his support,
encouragement, and excellent suggestions.  Thanks also to Scott Hughes  for
contributing through his ideas and enthusiasm for this project.  I would
especially like to thank Janna Levin for her careful reading of the paper and
insightful comments. Finally, I would like to thank Kip Thorne for teaching me
general relativity and Jim Yorke for teaching me dynamical systems theory.
\end{acknowledgments}

\appendix*
\section{Full Jacobian}
\label{app:jacobian}

For reference, we list the derivatives needed to calculate the full Jacobian
matrix.

From Sec.~\ref{sec:Jacobian_matrix}, we have the following:

\begin{eqnarray}
\frac{\partial \dot x^\mu}{\partial x^\nu}&=&N \left[p_\alpha g^{\alpha\mu}_{\ \ ,\nu} + 
	w^\mu_{\ ,\nu} 
\nonumber\right.\\
&&\left. +\, v^\mu(v_\alpha w^\alpha_{\ ,\nu} +
\textstyle{1\over2}N w^\alpha w^\beta g_{\alpha\beta,\nu})\right]
\end{eqnarray}

\begin{equation}
\frac{\partial \dot x^\mu}{\partial p_\nu}=N(g^{\mu\nu}+
	W^{\mu\nu}+Nv^\mu w^\nu)+Nv^\mu v_\alpha W^{\alpha\nu}
\end{equation}
with
\begin{equation}
W^{\mu\nu}=
-{^*}R^{*\mu\alpha\nu\beta}_{\ \ \ \ \ }S_\alpha S_\beta
\end{equation}

\begin{equation}
\frac{\partial \dot x^\mu}{\partial S_\nu}=NV^{\mu\nu}+Nv^\mu v_\alpha V^{\alpha\nu}
\end{equation}
with
\begin{equation}
V^{\mu\nu}=-S_\alpha p_\beta({^*}R^{*\mu\alpha\beta\nu}-
{^*}R^{*\mu\nu\alpha\beta})
\end{equation}

Now we simply apply the product rule many times:

\begin{eqnarray}
\frac{\partial \dot p_\mu}{\partial x^\nu}&=&
	-p_\alpha S_\beta\left(R^{*\ \ \alpha\beta}_{\mu\gamma\ \ \ ,\nu}\,v^\gamma +
	R^{*\ \ \alpha\beta}_{\mu\gamma}\,v^\gamma_{\ ,\nu}
\nonumber\right)\\
&&
+\, p_\alpha\left( \Gamma^\alpha_{\ \beta\mu,\nu}\,v^\beta +
	\Gamma^\alpha_{\ \beta\mu}\,v^\beta_{\ ,\nu}\right)
\end{eqnarray}

\begin{eqnarray}
\frac{\partial \dot p_\mu}{\partial p_\nu}&=&
	-S_\beta\left(R^{*\ \ \nu\beta}_{\mu\gamma\ \ }\,v^\gamma +
	R^{*\ \ \alpha\beta}_{\mu\gamma}\,p_\alpha
	\frac{\partial v^\gamma}{\partial p_\nu}
\nonumber\right)\\
&&
+\,  \Gamma^\nu_{\ \beta\mu}\,v^\beta +
	\Gamma^\alpha_{\ \beta\mu}\,p_\alpha\frac{\partial v^\beta}{\partial
	p_\nu}
\end{eqnarray}

\begin{eqnarray}
\frac{\partial \dot p_\mu}{\partial S_\nu}&=&
	-R^{*\ \ \alpha\nu}_{\mu\gamma\ \ }\,v^\gamma p_\alpha -
	R^{*\ \ \alpha\beta}_{\mu\gamma}\frac{\partial v^\gamma}{\partial S_\nu}
	p_\alpha S_\beta
\nonumber\\
&&
+\, \Gamma^\alpha_{\ \beta\mu}\,p_\alpha\frac{\partial v^\beta}{\partial S_\nu}
\end{eqnarray}

\begin{eqnarray}
\frac{\partial \dot S_\mu}{\partial x^\nu}&=&
	-p_\mu S_\alpha p_\gamma S_\delta
	\left(R^{*\alpha\ \gamma\delta}_{\ \ \beta\ \ \ ,\nu}\,v^\beta +
	R^{*\alpha\ \gamma\delta}_{\ \ \beta}\,v^\beta_{\ ,\nu}
\nonumber\right)\\
&&
+\, S_\alpha\left( \Gamma^\alpha_{\ \beta\mu,\nu}\,v^\beta +
	\Gamma^\alpha_{\ \beta\mu}\,v^\beta_{\ ,\nu}\right)
\end{eqnarray}

\begin{eqnarray}
\frac{\partial \dot S_\mu}{\partial p_\nu}&=&
	-S_\alpha S_\delta v^\beta
	\left(\delta_\mu^{\ \nu}R^{*\alpha\ \gamma\delta}_{\ \ \beta}\, p_\gamma\,+
	p_\mu R^{*\alpha\ \nu\delta}_{\ \ \beta}
\nonumber\right)\\
&&
	 -\, p_\mu R^{*\alpha\ \gamma\delta}_{\ \ \beta}S_\alpha
	\frac{\partial v^\beta}{\partial p_\nu} p_\gamma S_\delta
\nonumber\\
&&
+\, \Gamma^\alpha_{\ \beta\mu}\,\frac{\partial v^\beta}{\partial p_\nu} S_\alpha
\end{eqnarray}

\begin{eqnarray}
\frac{\partial \dot S_\mu}{\partial S_\nu}&=&
	-p_\mu p_\gamma v^\beta
	\left(R^{*\nu\ \gamma\delta}_{\ \ \beta}\, S_\delta +
	R^{*\alpha\ \gamma\nu}_{\ \ \beta}\, S_\alpha
\nonumber\right)\\
&&
	 -\, p_\mu R^{*\alpha\ \gamma\delta}_{\ \ \beta}S_\alpha
	\frac{\partial v^\beta}{\partial S_\nu} p_\gamma S_\delta
\nonumber\\
&&
+\, \Gamma^\nu_{\ \beta\mu}\,v^\beta
+ \Gamma^\alpha_{\ \beta\mu}\,\frac{\partial v^\beta}{\partial S_\nu} S_\alpha
\end{eqnarray}

Accidentally leaving off the final term  in 
$\displaystyle{\frac{\partial \dot S_\mu}{\partial S_\nu}}$ led to the robust but 
spurious chaotic behavior mentioned in Sec.~\ref{sec:numerical}.

\bibliography{mdh_2002_1}% uses BibTeX to make bibliography; see mdh_2002_1.bib

\end{document}